\documentclass[sigconf, 10pt, screen]{acmart}
\hypersetup{draft}

\renewcommand\footnotetextcopyrightpermission[1]{} 
\newif\ifdebugdoc\debugdocfalse
\newif\ifflow\flowfalse

\usepackage{hhline}
\usepackage[flushleft]{threeparttable}

\makeatletter
\def\subsubsection{\@startsection{subsubsection}{3}{10pt}%
                                 {-.5\baselineskip \@plus -2\p@ \@minus -.2\p@}%
                 {2.5\p@}{\@subsubsecfont}}
\makeatother

\ifdebugdoc
    \definecolor{ForestGreen}{rgb}{0.0, 0.27, 0.13}
    \newcommand{\outline}[1]{\textbf{\colorbox{yellow}{Outline:}\textcolor{red}{#1.}}}

    \newcommand{\add}[1]{\textcolor{red}{#1}}
    \newcommand{\del}[1]{\textcolor{blue}{\sout{#1}}}
    \newcommand{\todo}[1]{\textcolor{red}{{\bf TODO}: {#1}}}
    \newcommand{\save}[1]{}
    \newcommand{\yuanjie}[1]{\textcolor{red}{{\bf Yuanjie}: {#1}}}
     \newcommand{\lixin}[1]{\textcolor{blue}{{\bf Lixin}: {#1}}}
     \newcommand{\wei}[1]{\textcolor{cyan}{{\bf Wei}: {#1}}}
     \newcommand{\jiayi}[1]{\textcolor{violet}{{\bf Jiayi}: {#1}}}
      \newcommand{\guojie}[1]{\textcolor{orange}{{\bf Guojie}: {#1}}}
    \newcommand{\fixme}[1]{\textcolor{blue}{{\bf FIXME}: {#1}}}
    \newcommand{\question}[1]{\textcolor{purple}{{\bf Question}: {#1}}}
\else
    \newcommand{\outline}[1]{}

    \newcommand{\add}[1]{#1}
    \newcommand{\del}[1]{}
    \newcommand{\todo}[1]{}
    \newcommand{\save}[1]{}
    \newcommand{\yuanjie}[1]{}
    \newcommand{\lixin}[1]{}
    \newcommand{\jiayi}[1]{}
    \newcommand{\wei}[1]{}
    \newcommand{\guojie}[1]{}
    \newcommand{\fixme}[1]{}
	\newcommand{\question}[1]{}
\fi

\ifflow
    \newcommand{\p}[1]{\vskip 1ex\noindent\colorbox{yellow}{\parbox{\columnwidth}{\textbf{Point:} {#1}}}}
    \newcommand{\key}[1]{\vskip 1ex\noindent\colorbox{yellow}{\parbox{\columnwidth}{\textbf{Keys:} {#1}}}}
    \newcommand{\q}[1]{\vskip 1ex\noindent\colorbox{cyan}{\parbox{\columnwidth}{\textbf{Question:} {#1}}}}
\else
    \newcommand{\p}[1]{}
    \newcommand{\key}[1]{}
    \newcommand{\q}[1]{}
\fi

\newcommand{\paragraphb}[1]{\vspace{0mm}\noindent\textbf{#1}\quad}
\newcommand{\paragraphe}[1]{\vspace{0mm}\noindent\emph{#1}}

\def\name{\texttt{F-Rosette}\xspace}
\def\ie{i.e.\xspace}
\def\eg{e.g.\xspace}

\def\etc{etc\xspace}
\def\wrt{w.r.t.\xspace}

\usepackage[export]{adjustbox}
\usepackage{url}
\usepackage{subfig}
\usepackage{xcolor}
\usepackage{xspace}
\usepackage{amsmath}
\usepackage{balance}
\usepackage{booktabs}
\usepackage{tabularx}
\usepackage{multirow}
\usepackage{enumitem}
\usepackage{multirow}
\usepackage{graphicx}
\usepackage{amsfonts}
\usepackage{verbatim}
\usepackage{algorithm}
\usepackage{algpseudocode}
\makeatletter
\renewcommand{\ALG@beginalgorithmic}{\scriptsize}
\makeatother
\usepackage[normalem]{ulem}
\usepackage{paralist}
\usepackage{slashbox}
\usepackage{varwidth}
\usepackage{amsmath}

\algnewcommand\algorithmicswitch{\textbf{switch}}
\algnewcommand\algorithmiccase{\textbf{case}}
\algnewcommand\algorithmicassert{\texttt{assert}}
\algnewcommand\Assert[1]{\State \algorithmicassert(#1)}%
\algdef{SE}[SWITCH]{Switch}{EndSwitch}[1]{\algorithmicswitch\ #1\ \algorithmicdo}{\algorithmicend\ \algorithmicswitch}%
\algdef{SE}[CASE]{Case}{EndCase}[1]{\algorithmiccase\ #1}{\algorithmicend\ \algorithmiccase}%
\algtext*{EndSwitch}%
\algtext*{EndCase}%

\usepackage{siunitx}
\usepackage{soul}
\usepackage{tikz}
\usetikzlibrary{arrows,chains,matrix,positioning,scopes}
\makeatletter
\tikzset{join/.code=\tikzset{after node path={%
\ifx\tikzchainprevious\pgfutil@empty\else(\tikzchainprevious)%
edge[every join]#1(\tikzchaincurrent)\fi}}}
\makeatother
\tikzset{>=stealth',every on chain/.append style={join},
         every join/.style={->}}
\tikzstyle{labeled}=[execute at begin node=$\scriptstyle,
   execute at end node=$]
\tikzset{skip loop/.style={to path={-- ++(0,#1) -| (\tikztotarget)}},
         hv path/.style={to path={-| (\tikztotarget)}},
         vh path/.style={to path={|- (\tikztotarget)}}}

\newcolumntype{L}[1]{>{\raggedright\let\newline\\\arraybackslash\hspace{-2pt}}m{#1}}
\newcolumntype{C}[1]{>{\centering\let\newline\\\arraybackslash\hspace{-2pt}}m{#1}}
\newcolumntype{R}[1]{>{\raggedleft\let\newline\\\arraybackslash\hspace{-2pt}}m{#1}}

\theoremstyle{plain}
  \newtheorem{thm}{Theorem}
  \newtheorem{prop}{Property}

\theoremstyle{acmdefinition}
  
  \newtheorem{lem}{Lemma}

\theoremstyle{remark}

\usepackage{mathtools}

\usepackage[skip=6pt]{caption}
\usepackage{xr}

\usepackage{xcolor}

\usepackage{tablefootnote}

\PassOptionsToPackage{hyphens}{url}\usepackage{hyperref}

\begin{document}

\title{\huge Fractal Rosette:\\A Stable Space-Ground Network Structure in Mega-constellation}

\iftrue
\author{
Yuanjie Li, Hewu Li, Lixin Liu, Wei Liu, Jiayi Liu, Jianping Wu,
}
\author{Qian Wu, Jun Liu, Zeqi Lai, Guojie Fan}
\affiliation{%
   \institution{Institute for Network Sciences and Cyberspace, Tsinghua University}
 }
\fi

\begin{abstract}

We present \name, a stable space-ground network structure for low-earth orbit (LEO) satellite mega-constellations at scale. 
Due to the dynamic many-to-many space-ground mapping in high mobility, existing LEO mega-constellations with IP protocol stack suffer from 
frequent user IP address changes (every 133--510s per user) and network routing re-convergence (thus $\leq$20\% network usability). 
To provably stabilize the space-ground network under high dynamics,
\name aligns the network design in the cyberspace with the mega-constellations in the physical world.
It devises a recursively stable network structure over the Rosette constellation, 
decouples the hierarchical network addresses from mobility for stability,
and aligns the geographical routing on the ground with the topological routing in space for efficiency and high usability. 
Our hardware-in-the-loop, trace-driven emulations validate \name's stability, near-optimal routing ($\leq$1.4\% additional delays), and affordable overhead (<1\% CPU,  <2MB memory) for resource-constrained satellites.

\end{abstract}


\maketitle
\pagestyle{plain}

\section{Introduction}
\label{sect:intro}

The future Internet is up in the sky. 
Since 2019, we have witnessed a rocket-fast deployment of the low-earth-orbit (LEO) satellite mega-constellations, including SpaceX Starlink \cite{starlink}, Amazon Kuiper \cite{kuiper}, Telesat \cite{telesat}, OneWeb \cite{oneweb}, to name a few. 
Compared to the traditional satellite networks, 
these LEO mega-constellations promise ultra-low network latency and high throughput that are competitive to the terrestrial networks. 
Moreover, with massive satellites, these mega-constellations offer the global Internet coverage to users that were not easily reachable by terrestrial networks, such as those in rural areas and the ocean.

This work asks a simple question: {\em How to enable a usable network in mega-constellations to serve the terrestrial users?}
Despite the rapid deployments of LEO mega-constellations, inter-connecting satellites for network services is still at early designs on a tiny scale (such as Starlink's ``Better Than Nothing'' plan \cite{starlink-data-plan} and initial inter-satellite link test \cite{starlink-isl-test,starlink-isl-test-2}). 
With massive mobile LEO satellites and their relative motions to numerous terrestrial users on a rotating earth, 
it is open to question whether the traditional wisdom of satellite network designs are still suitable for the mega-constellations. 

A grand challenge for a usable space-ground network is to guarantee its {\em stability} under satellites and earth's high mobility.
Unlike the traditional satellite and terrestrial networks, the LEO network exhibits {\em highly dynamic many-to-many mapping} between the moving satellites and terrestrial users.
Each LEO satellite can cover $\approx$100,000 users,
while each terrestrial user has visibility to $\approx$20 satellites in mega-constellations \cite{del2019technical}. 
The high satellite mobility and earth's rotation force the terrestrial users to frequently re-associate to new satellites.
Therefore, the network topology between the space and ground repetitively changes, which results in frequent network address changes and routing protocol re-convergence.
Our projection of mega-constellations running IP protocols shows that ($\S$\ref{sec:challenge}), every terrestrial user switches its IP address every 133--510s (with 2,082--7,961 users changing IP addresses every second), and the network usability decreases to $\leq$20\% with frequent routing re-convergence.
Even worse, both stability issues are amplified as {mega}-constellations scale to massive satellites.

The fundamental cause of the above unstable space-ground network is the mismatch between the network in the {virtual cyberspace} and mega-constellations in the {physical world}.
On one hand, classical mega-constellations primarily focus on offering good coverage, but neglects the demands of stable network topology, addressing, and routing. 
On the other hand, most satellite networks in the virtual cyberspace heavily rely on the logical address and routing, which are vulnerable to the high satellite mobility earth's rotations in reality.
Therefore, a stable space-ground network at scale calls for a coherent design between cyberspace and reality.

We propose \name, 
a stable space-ground network structure in LEO mega-constellation. 
\name stabilizes the space-ground network in the virtual cyberspace by aligning it with the dynamic mega-constellations in the physical world.
With the Rosette constellation as the basic building block, \name recursively constructs a provably stable network topology with inter-satellite links. 
On top of this stable network topology, we derive a hierarchical network address space for the satellites and terrestrial users.
To prevent frequent address changes, we decouple this address space from mobility in \name's new geographical coordinate system. 
Routing with \name's network address is stable and efficient {\em without} requiring routing re-convergence, thus retaining high network usability.
Routing between satellites follows the standard IPv6 prefix/wildcard matching with provable optimality and multi-path support. 
To route traffic between terrestrial users, \name performs local geographical routing by embedding it into the topological routing between satellites over the stable network topology.
For practical deployment, \name is backward compatible with IPv6, supports incremental expansion, and facilitates inter-networking to external networks. 

We prototype \name's protocol suite and evaluate it with hardware-in-the-loop, trace-driven emulations.
Compared to existing mega-constellations, 
\name stabilizes network topology, addressing, and routing with near-optimal data delivery ($\leq$1.4\% additional delay). 
\name incurs negligible CPU (<1\%) and memory (<2MB) cost that is comparable to standard IPv6, thus suitable for small LEO satellites.
\section{Space-Ground Network Primer}
\label{sec:back}


\todo{Unify and simplify the notations. Define and minimize the necessary terms}

This section introduces the LEO satellite mega-constellations today ($\S$\ref{sec:back:back}),
and analyzes why they cannot guarantee stable space-ground network services ($\S$\ref{sec:challenge}). 

\subsection{Satellites and Mega-Constellations}
\label{sec:back:back}


\begin{figure}[t]
\vspace{-10mm}
\includegraphics[width=.7\columnwidth]{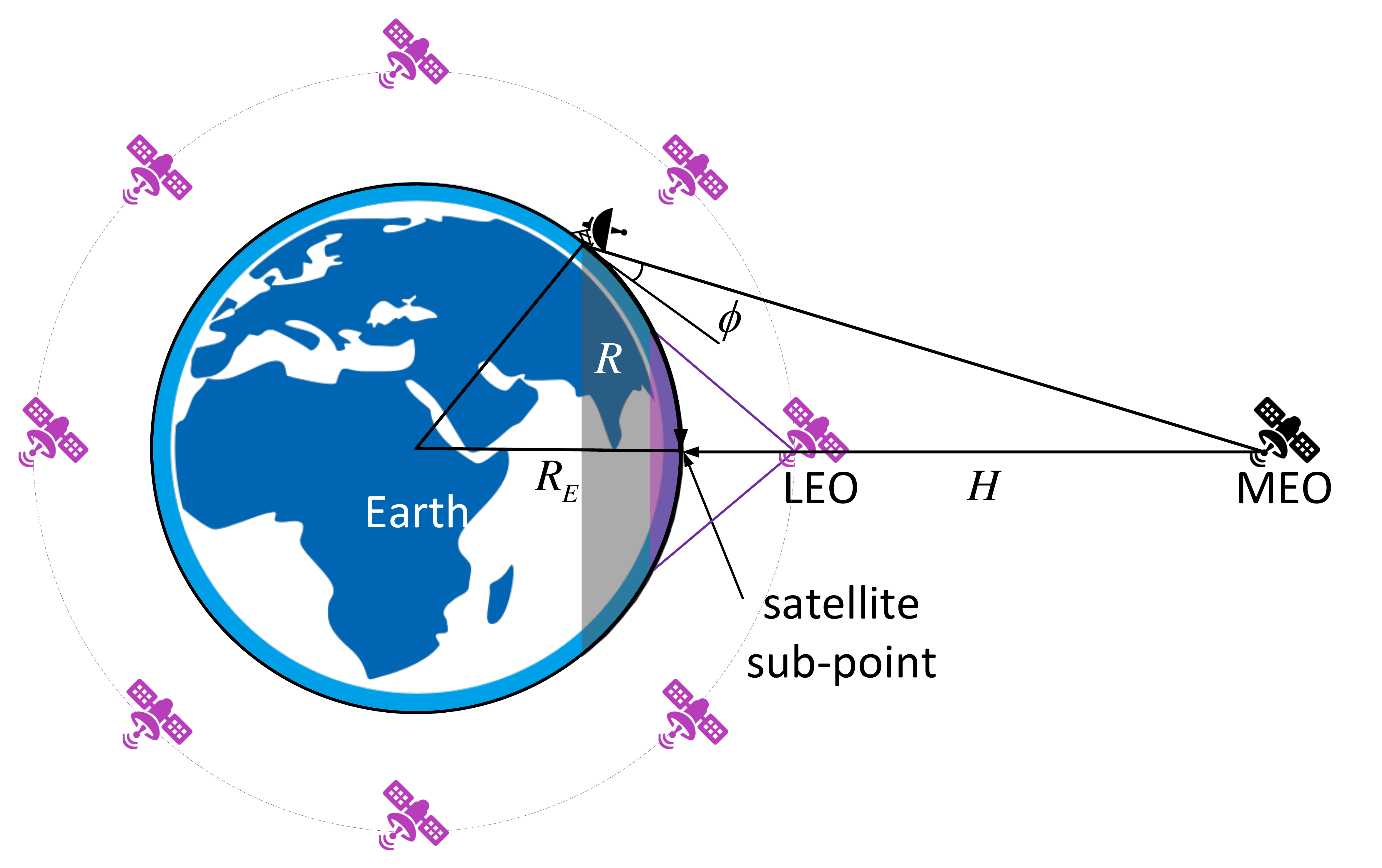}
\vspace{-2mm}
\caption{
Satellites, constellations, and their coverages.
}
\label{fig:coverage}
\end{figure}
\begin{table}[t]
\vspace{-2mm}
\resizebox{1\columnwidth}{!}{{
\begin{tabular}{|c|c|c|c|c|c|}
\hline
& {\bf Num.} & {\bf Num.} & {\bf Altitude} & {\bf Inclination} & {\bf Ground-to-}\\
& {\bf satellites} & {\bf orbits} & {\bf $H$ (km)} & {\bf angle $\phi$} & {\bf space RTT (ms)}\\
\hline
{\bf Starlink~\cite{starlink}} &\begin{tabular}[c]{@{}l@{}}1584/1584/\\ 720/348/172\end{tabular} &\begin{tabular}[c]{@{}l@{}}72/72/\\ 36/6/4\end{tabular} &\begin{tabular}[c]{@{}l@{}}550/540/\\ 570/560/560\end{tabular} &\begin{tabular}[c]{@{}l@{}}53/53/\\ 70/97/97\end{tabular}&\begin{tabular}[c]{@{}l@{}}3.7/3.6/\\ 3.8/3.7/3.7\end{tabular}\\
\hline
{\bf Kuiper~\cite{kuiper}} & 1156/1296/784 &34/36/28 & 630/610/590 & 52/42/33&4.2/4.07/3.93\\
\hline
	{\bf Telesat~\cite{telesat}}&351/1320 &27/40& 1015/1325&98.98/50.88 & 6.77/8.83\\
\hline
{\bf Iridium~\cite{tan2019new}} &66&6& 725&86.4& 4.83\\
\hline
\end{tabular}
}}
\vspace{1mm}
\caption{Statistics of common LEO constellations. 
}
\label{tab:mega-constellation}
\end{table}

Satellites can be classified by their orbits' altitude. 
A satellite can operate at low-earth orbit (LEO, $\leq$2,000 km), medium-earth orbit (MEO, $\leq$35,786 km), or high-earth orbit (HEO, >35,786 km).
An orbit can be described by its altitude, {inclination} to Equator, and {right ascension angle}\footnote{In theory, 7 parameters are required to describe an orbit \cite{orbit-element}. But real constellations need fewer degrees, \eg, circular orbit with zero eccentricity and same apogee/perigee. So we will use a simpler version in this paper.}.
Traditional communication satellites usually operate at MEO orbits, such as the geostationary orbit (GEO, 35,786 km). 
As shown in Figure~\ref{fig:coverage}, a higher-altitude satellite offers larger coverage (defined as its great circle range $R$ on the earth) but at the cost of longer ground-to-space round-trip latency.

A small LEO satellite usually has very limited coverage, computation power, and storage.
To enable ubiquitous network services, one possible solution is to deploy an LEO {\em mega-constellation}. 
Table~\ref{tab:mega-constellation} summarizes the statistics of common mega-constellations today.
The massive satellites help fully cover the areas of interest, while the LEO orbits ensure low ground-to-space round trips.
Each satellite uses its microwave radio link for data transfer from/to terrestrial users. 
Satellites can interconnect with each other via free-space laser inter-satellite links. 
Unlike the traditional communications satellites at link layer only, LEO satellites in the mega-constellations should route traffic at the network layer and inter-connect to other terrestrial networks (\eg, via TCP/IP) for {\em global} Internet services {\em at scale} (acknowledged by Elon Musk \cite{starlink-ipv6} and recent work \cite{handley2018delay,bhattacherjee2019network}).
\subsection{Unstable Space-Ground Network Today}
\label{sec:challenge}



The LEO mega-constellations aim to enable usable and performant network services for terrestrial users.
Unfortunately, our analysis and trace-driven emulation project that, {with the {\em dynamic many-to-many} space-ground mapping in high mobility, existing mega-constellations running standard IP will suffer from frequent IP address changes for the users and routing re-convergence for the network (thus low usability).}

\paragraphb{Why unstable: Dynamic many-to-many mapping}
In LEO mega-constellations, the many-to-many mapping between satellites and terrestrial users is a norm rather than an exception.
Each LEO satellite can cover many users ($\approx$100,000), \jiayi{Check the number}
while each terrestrial user has visibility to many satellites ($\approx$20) in mega-constellations. \jiayi{Check the number}
This many-to-many mapping is dynamic due to high mobility.
Compared to the MEO/GEO satellites, LEO satellites at a much lower altitude (typically $\leq$2,000 km) exhibit high mobility\footnote{According to Kepler's third law, for all satellites of the earth, the ratio between their orbital altitude and period $(R_E+H_s)^3/T_s^2$ is a constant.} and relative motions.
Moreover, terrestrial users also exhibit relative motions to the satellites due to the earth's rotations.
Both result in frequent changes of the network topology between space and ground.

\begin{figure}[t]
\vspace{-5mm}
\subfloat[Interval between changes]{
\includegraphics[width=0.5\columnwidth]{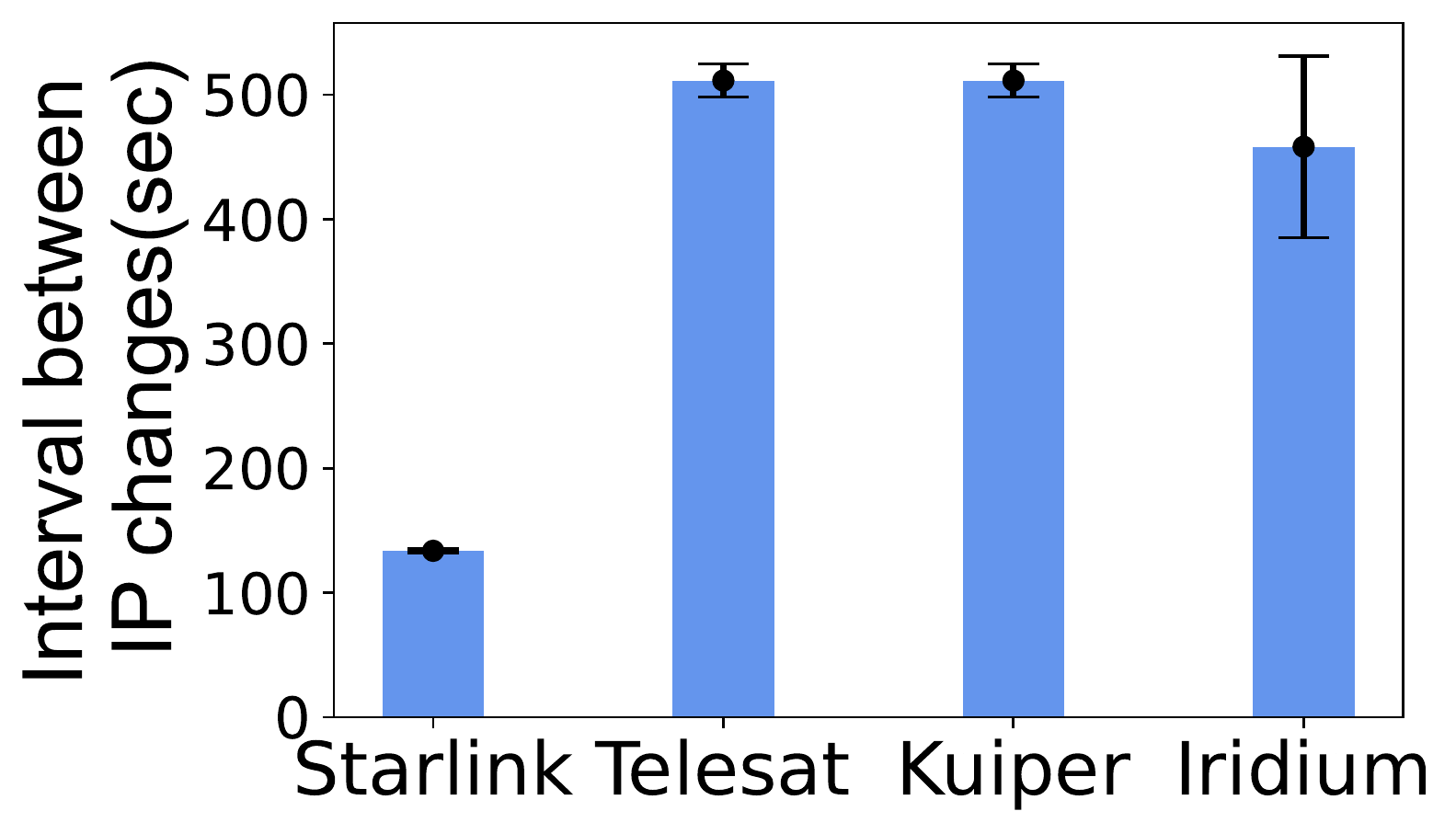}
\label{fig:user_ip_change}
}
\subfloat[IP changed users per second]{
\includegraphics[width=0.5\columnwidth]{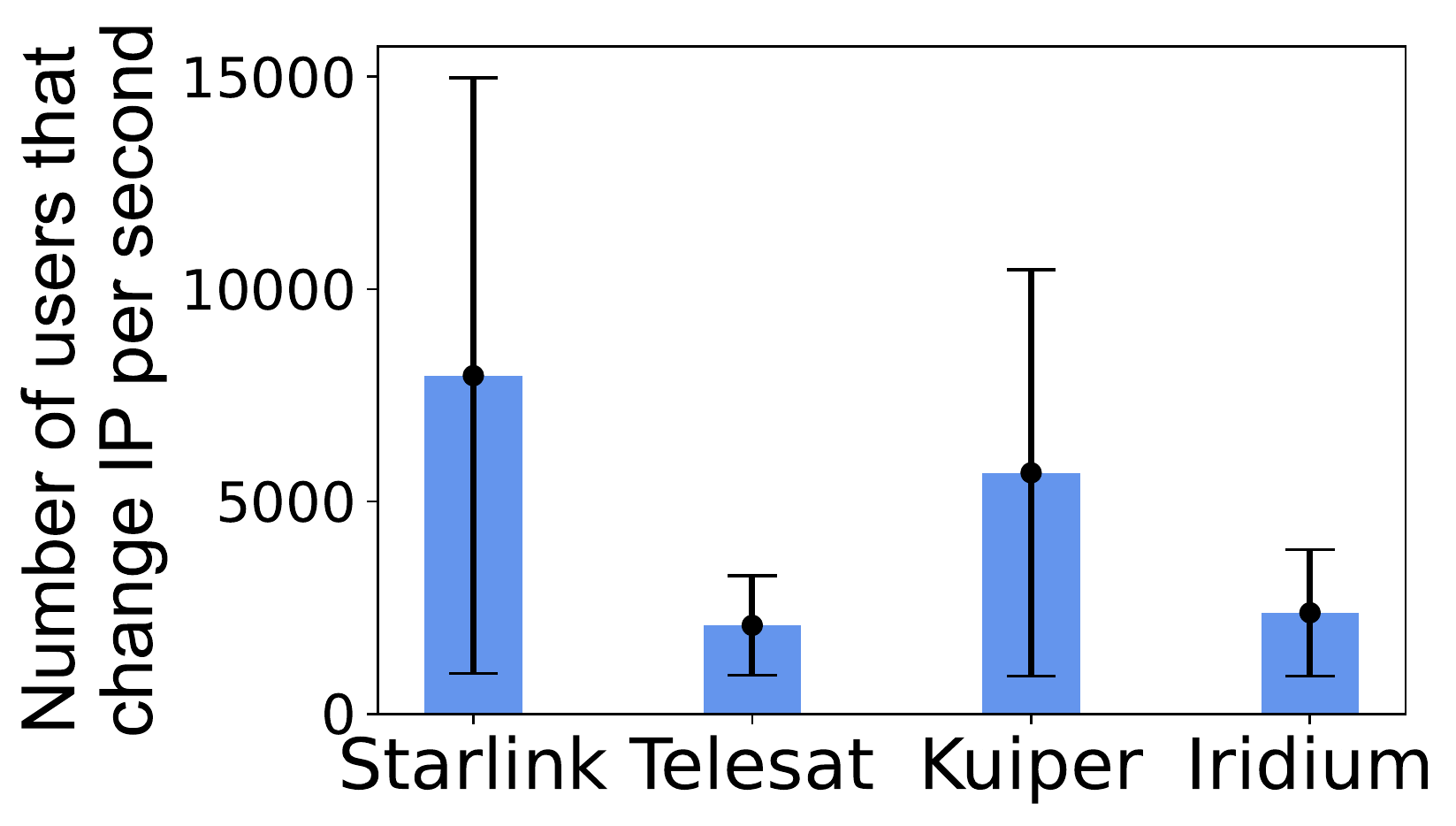}
\label{fig:timecdf_ip_change}
}
\vspace{-2mm}
\caption{Projected users' IP address change frequency in LEO mega-constellations in the direct access mode.}
\label{fig:ip_change}
\end{figure}

\begin{figure}[t]
\vspace{-5mm}
\subfloat[Starlink]{
\includegraphics[width=0.5\columnwidth]{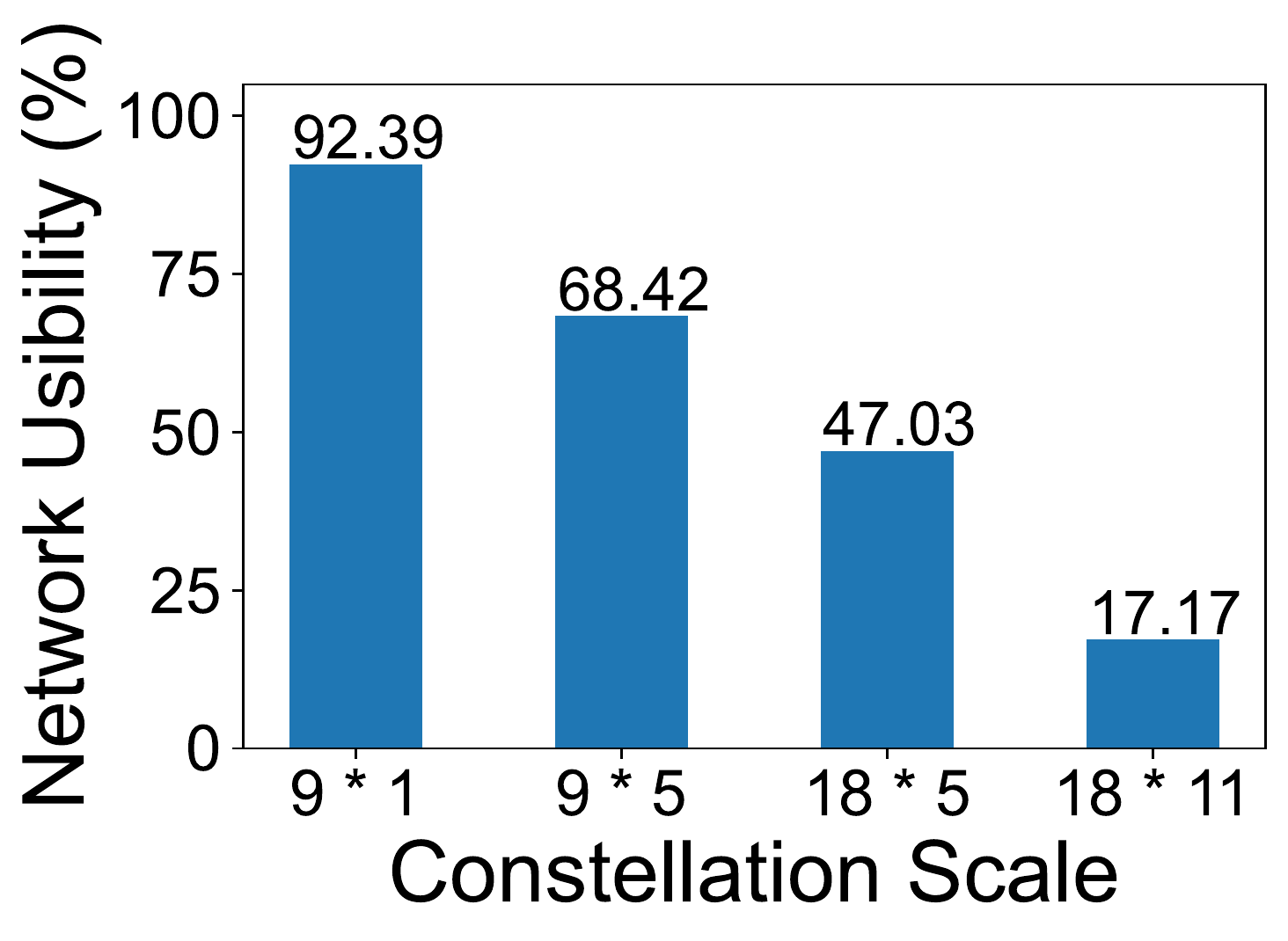}
\label{fig:starlink_usibility}
}
\subfloat[Kuiper]{
\includegraphics[width=0.5\columnwidth]{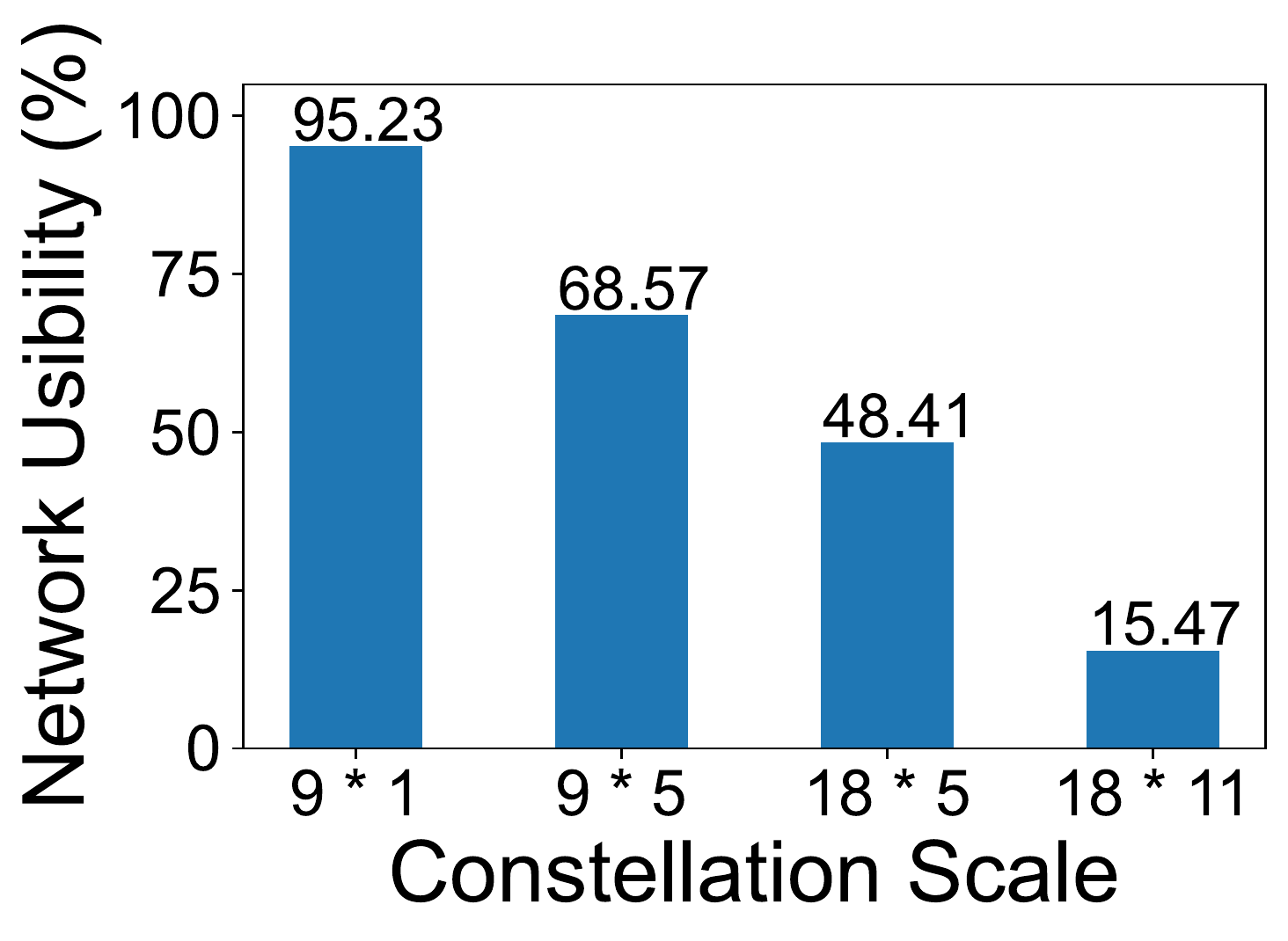}
\label{fig:kuiper_usibility}
}
\vspace{-2mm}
\caption{Projected network usability in LEO mega-constellations in ground station-assisted mode.}
\label{fig:usability}
\end{figure}

\paragraphb{How (un)stable is the space-ground network:}
We next quantify the impact of dynamic many-to-many mapping on the stability of the space-ground network.
Since the LEO mega-constellations are still at their early deployments, their network designs are still underway and unfinished \cite{starlink-data-plan, starlink-isl-test,starlink-isl-test-2}.
To this end, we conduct a trace-driven projection based on the operational mega-constellations in Table~\ref{tab:mega-constellation} over IPv6 in a hardware-in-the-loop emulator (detailed in $\S$\ref{sect:eval}). 
We consider two usage modes:
(1) {\em Direct access:} A terrestrial user with a satellite phone directly accesses the satellite network.
We follow the world population statistics from NASA in 2020 \cite{NASA2020} and Starlink's estimated customer numbers in the U.S. \cite{starlink-users} to proportionally estimate the users that directly access satellite network at different locations.
(2) {\em Ground station-assisted access:} Terrestrial users associate to a ground station \cite{aws-gs,ms-gs}, which connects to a satellite for network access.
We assume six ground stations in New York, Tokyo, Beijing, Hong Kong, Shanghai, and Singapore; the results for ground stations at other locations are similar.
We assume each user or ground station will always associate to the physically nearest satellite for good coverage \cite{chowdhury2006handover}.
All network nodes run the IPv6 stack and OSPF.
Figure~\ref{fig:ip_change} and Figure \ref{fig:usability} show the projected impact on the users and the network, respectively.

{\em $\circ$ Impact on users: Frequent IP address changes.}
In the direct access mode, a terrestrial user's IP address belongs to the subnet of the corresponding satellite logical network interface.
With rapid satellite motions, the terrestrial user has to frequently handoff to new satellites (thus new network interfaces and addresses) to retain its Internet access. 
{As shown in Figure~\ref{fig:ip_change}, each user is forced to change its logical IP address every 133--510s, and every second we observe \add{2082--7961} global users per second should change their IP addresses.}
Such frequent address changes will repetitively disrupt the TCP connections and upper-layer applications, and will negatively impact the user experiences. 


{\em $\circ$ Impact on network: Frequent routing re-convergence and low usability.}
To prevent frequent user IP changes, one can adopt the ground station-assisted access mode. 
In this mode, the ground station is a gateway for a fixed IP subnet, so that all users associating with it retain the time-invariant address.
As the ground station re-associates to a new satellite, the space-ground logical network topology changes.
Then the IP routing must re-converge to guarantee reachability to this ground station, before which the network cannot route traffic to the ground station.
Note each LEO satellite can only provide very short coverage for a ground station (10 minutes in Iridium, $\leq$3 minutes in Starlink).
Frequent handoffs between satellites result in repetitive routing re-convergence and thus low network usability (\ie, $1-\frac{\text{routing re-convergence duration}}{\text{each satellite's service duration}}$).
Figure~\ref{fig:usability} shows that, all mega-constellations today suffer from $\leq20\%$ network usability even under small constellations.
The network usability decreases with more satellites, which causes more frequent handoffs for ground stations.

\paragraphb{How mega-constellation amplifies instability:}
Both issues will worsen with more satellites in the mega-constellations.
Users in the direct access mode will experience more handoffs with more satellites, thus causing more frequent IP address changes.
For the network in the ground station-assisted mode, Figure~\ref{fig:usability} shows more satellites prolong the routing re-convergence on link change, thus lowering the network usability. 
Moreover, more satellites complicate the dynamic routing between the moving satellites and terrestrial users (due to the earth's rotation). 
Solutions for small LEO networks decades ago (\eg, virtual topology \cite{werner1997dynamic, chang1998fsa} and virtual nodes \cite{ekici2000datagram, ekici2001distributed}) cannot scale to the mega-constellations today due to their vulnerability to dynamic many-to-many mapping, and expensive (pre-)computation and storage for small LEO satellites. 
Motivated by this, we next design \name, a stable space-ground network structure for LEO mega-constellations.

\section{The \name Network Structure}
\label{sec:arch}

We introduce \name's structure and its basic properties. 

\subsection{A Primer of Rosette Constellation}
\label{sec:rosette-primer}

The basic building block of \name is a Rosette constellation, which was originally proposed in \cite{ballard1980rosette} to guarantee full coverage with as few satellites as possible.
This section introduces the necessary background of Rosette constellation for \name; a complete description is available in \cite{ballard1980rosette}.

Figure~\ref{fig:rosette-example} exemplifies the Rosette constellation with 8 satellites. 
Each orbit has one satellite, and all satellites move in circular orbits 
at the altitude of $H$ (thus period $T$).
A Rosette constellation is defined as a tuple $(N, m)$, where $N$ is the number of satellites
and $m\in\mathbb{N}$ is a harmonic phase shift.
All satellites are numbered from 0 to $N-1$.
For satellite $i$, its runtime location can be described by three constant angles $(\alpha_i,\beta_i,\gamma_i)$ and one time-varying phase angle $x_t$ at time $t$:


\begin{align}
\alpha_i &= \frac{2\pi i}{N}, i=0,1,...,N-1 \label{eqn:alpha}\\
\beta_i &\equiv \beta, \forall i \label{eqn:beta}\\
\gamma_i &= m\alpha_i=m\cdot\frac{2\pi i}{N} \label{eqn:gamma}\\
x_t &= \frac{2\pi}{T_s}t \label{eqn:x}
\end{align}
where $\alpha_i$ is satellite $i$'s right ascension angle, $\beta_i\equiv\beta$ is the inclination angle for all orbits, $\gamma_i$ is $i$'s initial phase angle in its orbit at time $t=0$, and $x_t$ is the time-varying phase angle ({\em identical for all satellites anytime}). 
Compared to other constellations, the Rosette constellation has two appealing properties for networking:

\begin{figure}[t]
\vspace{-10mm}
\subfloat[Physical constellation in space]{
\includegraphics[width=0.51\columnwidth]{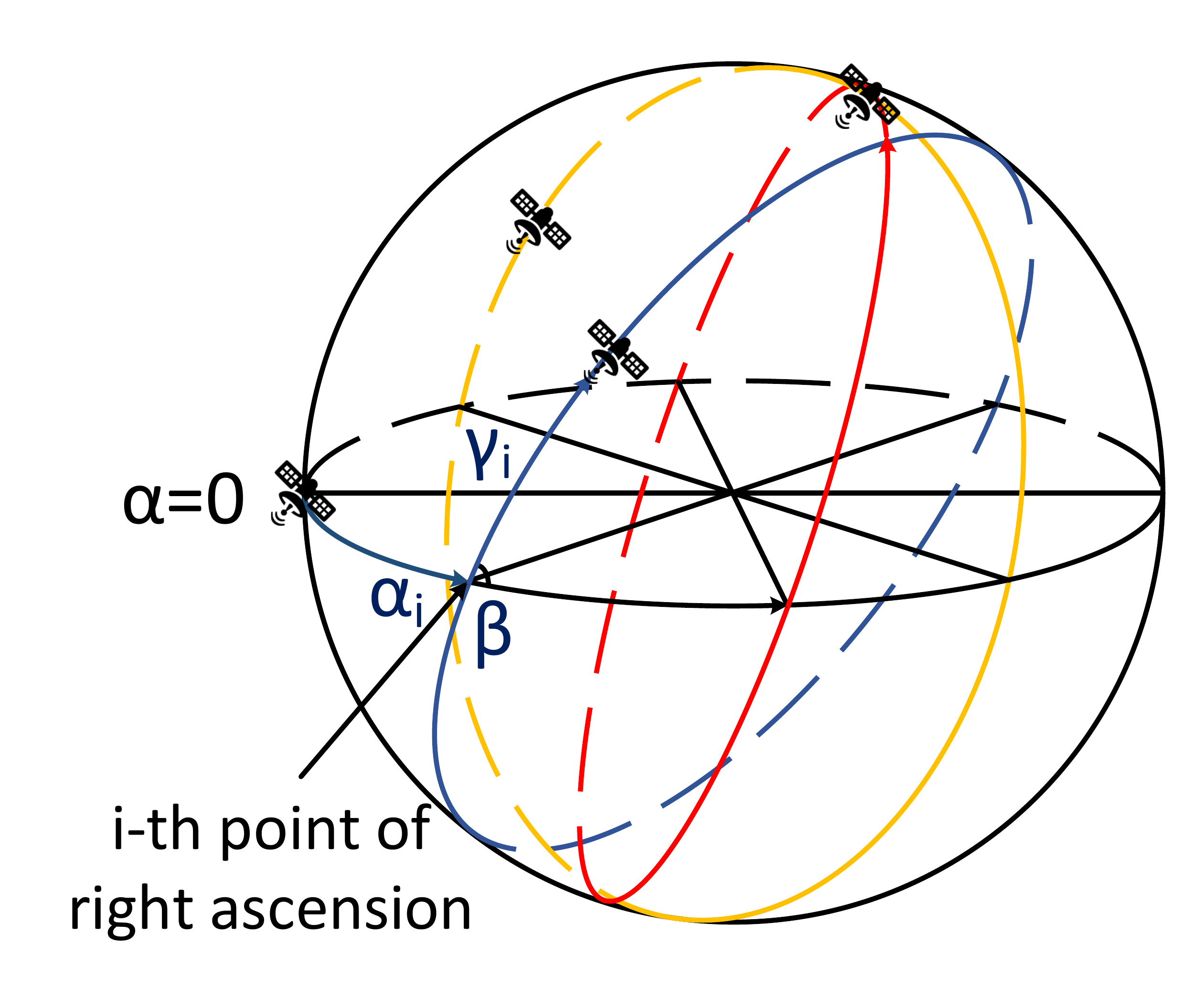}
\label{fig:rosette-constellation}
}
\subfloat[Logical topology in $\name_0$ \todo{Clarify local connection is {\em NOT} the only choice for \name.}]{
\includegraphics[width=0.49\columnwidth]{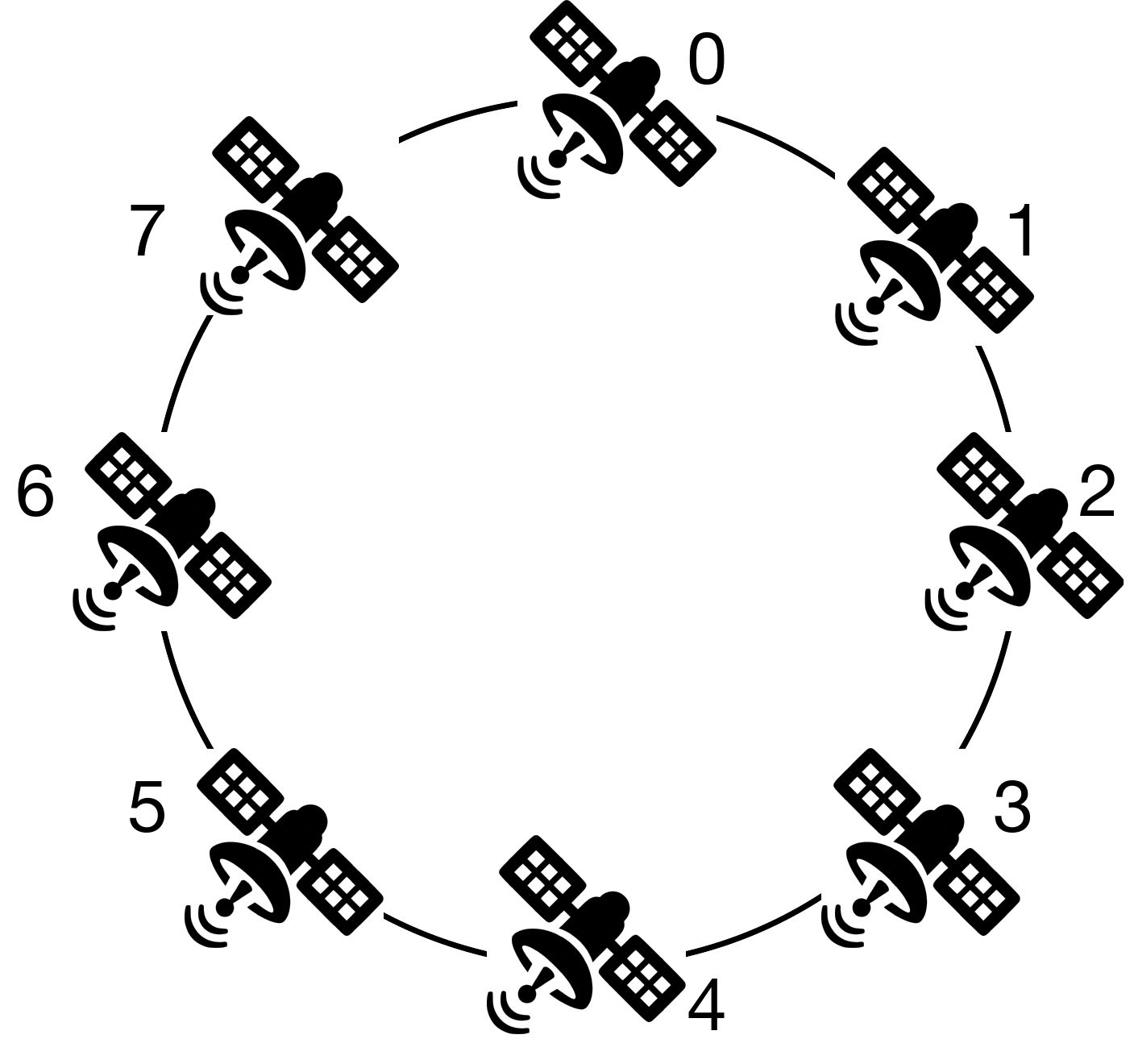}
\label{fig:base-case}
}
\vspace{-2mm}
\caption{A Rosette constellation with $N=8,m=6$.}
\label{fig:rosette-example}
\end{figure}

\paragraphb{Full coverage with fewer satellites:} 
The Rosette constellation usually needs fewer satellites than Walker (building block for Starlink and Kuiper) and polar orbits for full coverage (see \cite{ballard1980rosette} and $\S$\ref{sec:eval:characteristics} for the evaluations). 
Each satellite's coverage $R$  in Figure~\ref{fig:coverage} can be determined by its altitude $H$ and elevation angle $\phi$:
\begin{equation}
\tan\phi=\frac{\cos R-R_E/(R_E+H)}{\sin R}
\label{eqn:altitude}
\end{equation}

For LEO satellites, given each satellite's coverage $R$ on the earth, the minimum number of satellites $N_{min}$ needed to fully cover the earth is decided by the following relation:

\begin{equation}
\sec R=\sqrt{3}\tan{\left(\frac{\pi}{6}\frac{N_{min}}{N_{min}-2}\right)}
\label{eqn:per-sate-coverage}
\end{equation}

\paragraphb{Periodic ground-track repeat orbits:} 
Due to the earth rotation and high LEO satellite mobility, the satellite sub-point in Figure~\ref{fig:coverage} is time-varying and non-repeatable in general\footnote{The geostationary satellites guarantee time-invariant satellite sub-point, but at the altitude of 35,786km and thus suffer from long network latency.} (exemplified in Figure~\ref{fig:iterative-expansion}). 
This complicates locating and routing between the ground and space. 
Instead, the Rosette constellation allows LEO satellites to retain periodic satellite sub-points via ground-track repeat orbits, \ie, every satellite will {\em periodically} revisit its previous satellite sub-point. 
To do so, we configure the satellites' orbital period $T_s$ such that 
\begin{equation}
T=T_E/(N-m)
\label{eqn:repeat-orbit}
\end{equation}
where $T_E$ is the period of earth rotation. For each satellite $i$, the latitude $\varphi_i$ and longitude $\lambda_i$ of its sub-point is: 
\begin{align}
&\sin\varphi_i = \sin\beta\sin{m(\omega_Et-\alpha_i)} \label{eqn:phi}\\
&\tan{\left[\lambda_i+m(\omega_Et-\alpha_i)\right]}=\cos\beta\tan m(\omega_Et-\alpha_i) \label{eqn:lambda}
\end{align}
where $\omega_E=2\pi/T_E$ is the earth's angular rotation speed.

\medskip
Despite these merits, however, the original Rosette constellation is not immediately applicable to networking. 
It was primarily designed for full coverage rather than network routing. 
There was no specification on how to inter-connect satellites, how to forward traffic among satellites, and how to route data among terrestrial users via satellites. 
This motivates us to re-struct the Rosette constellation for stable space-ground network at scale.

\subsection{\name Construction}
\label{sec:construction}

\begin{figure*}[t]
\vspace{-18mm}
\centering
\subfloat[\textsf{Shift(\name, t)} in Algorithm~\ref{algo:construction}]{
\includegraphics[width=.3\textwidth]{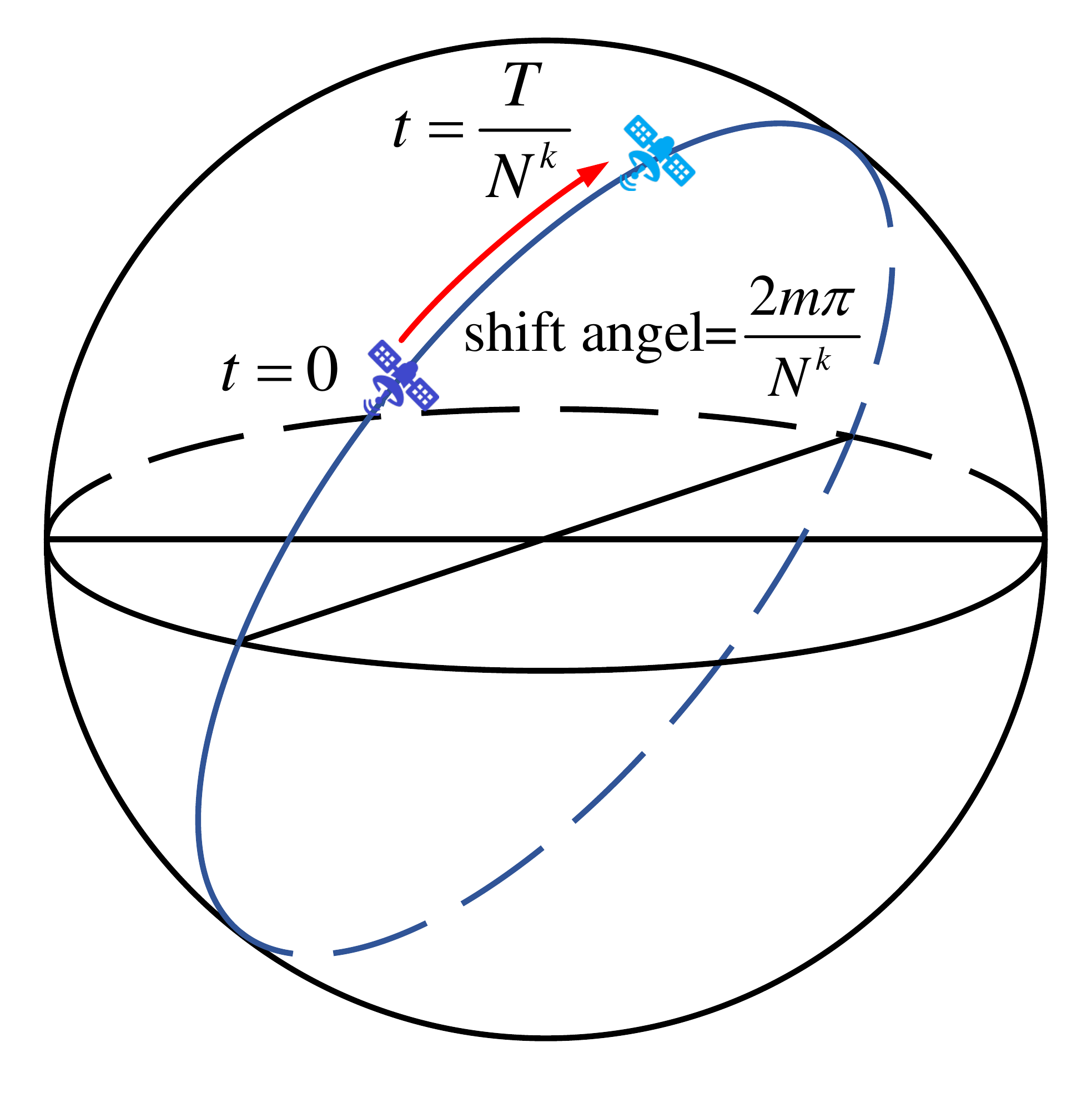}
\label{fig:shift}
}
\subfloat[Logical network topology of $\name_k$]{
\includegraphics[width=.6\textwidth]{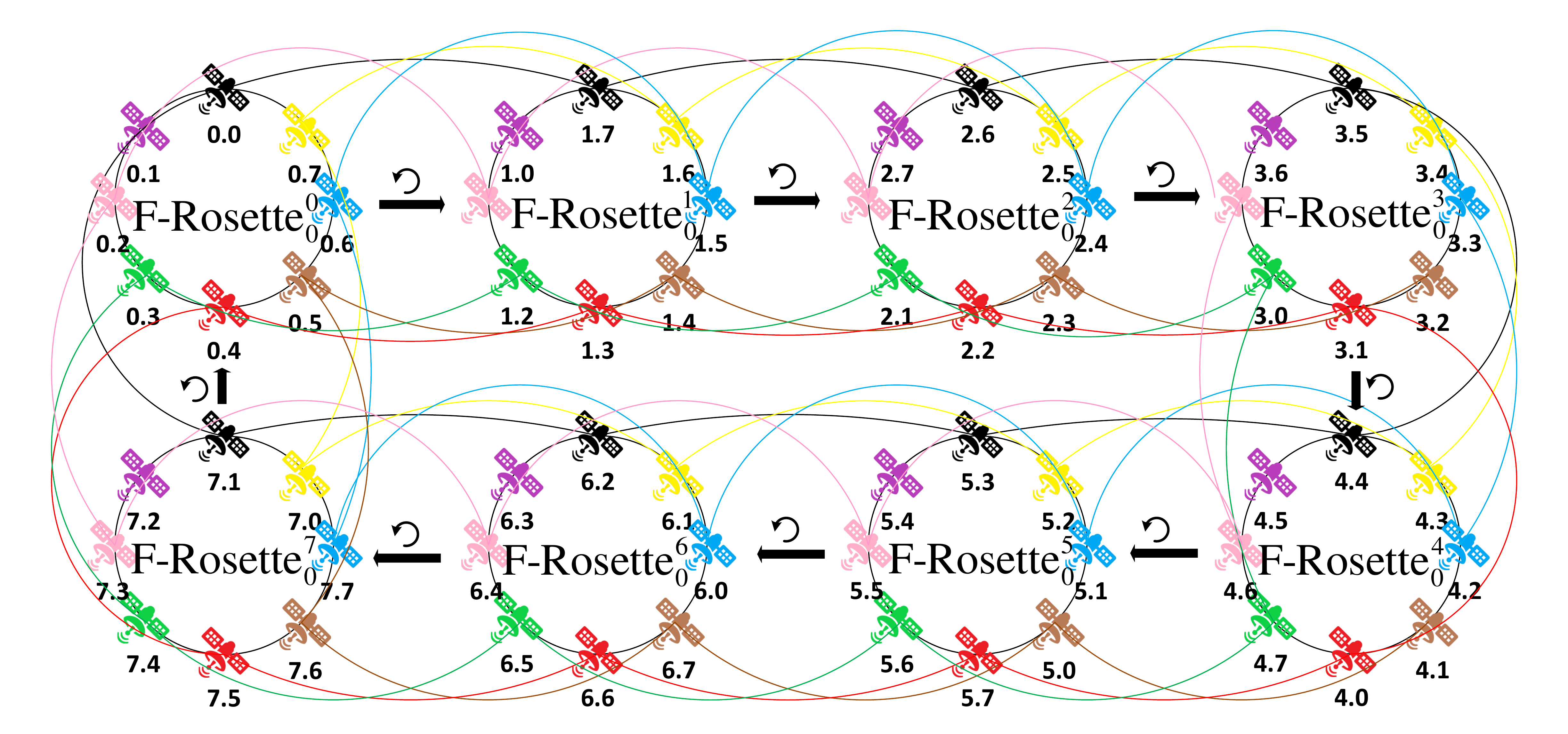}
\label{fig:logic-topology}
}
\vspace{-3mm}
\subfloat[Physical constellation in 3D space]{
\includegraphics[width=.3\textwidth]{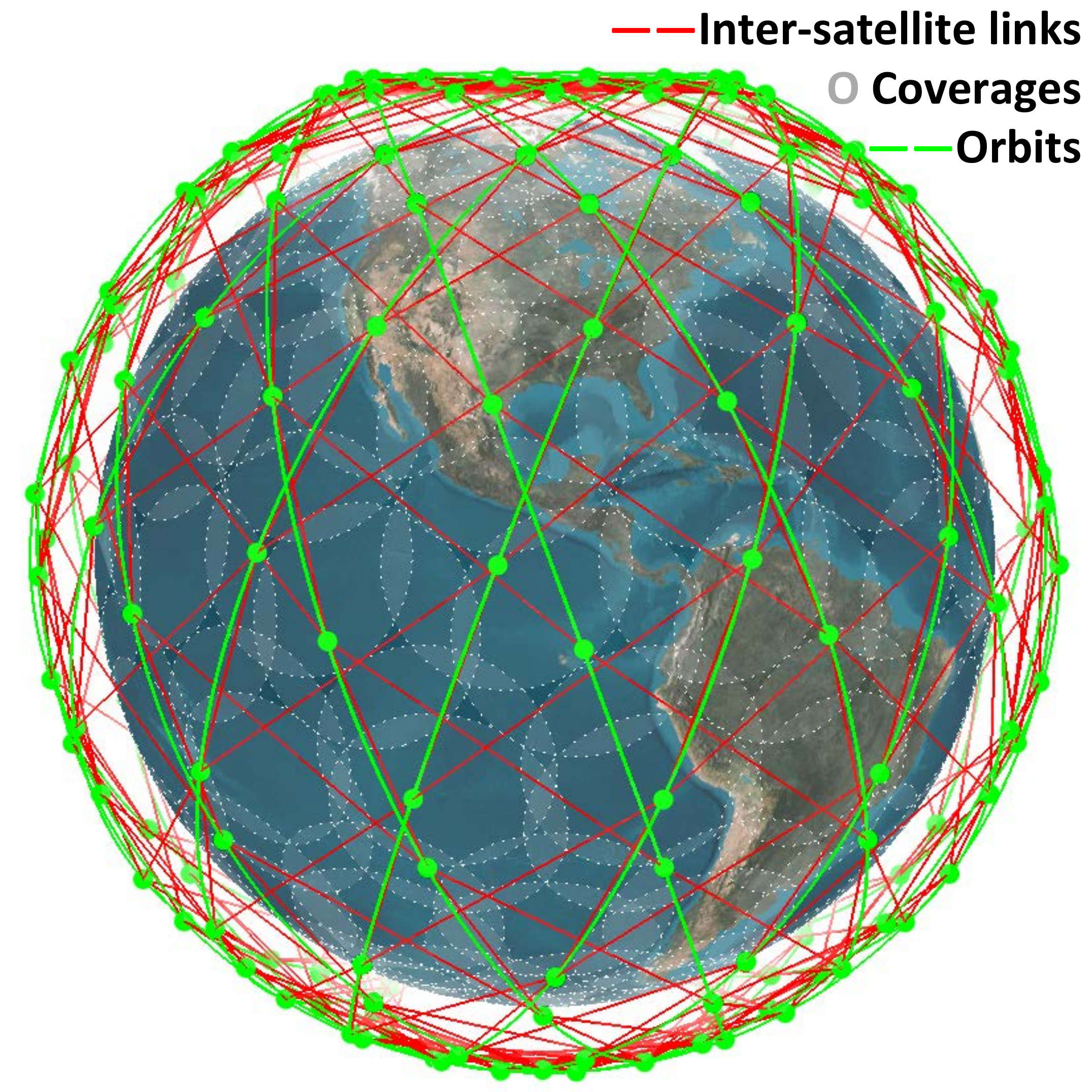}
\label{fig:3d-constellation}
}
\subfloat[Hierarchical geographical cells by fixed satellite sub-point trajectories.]{
\includegraphics[width=.6\textwidth]{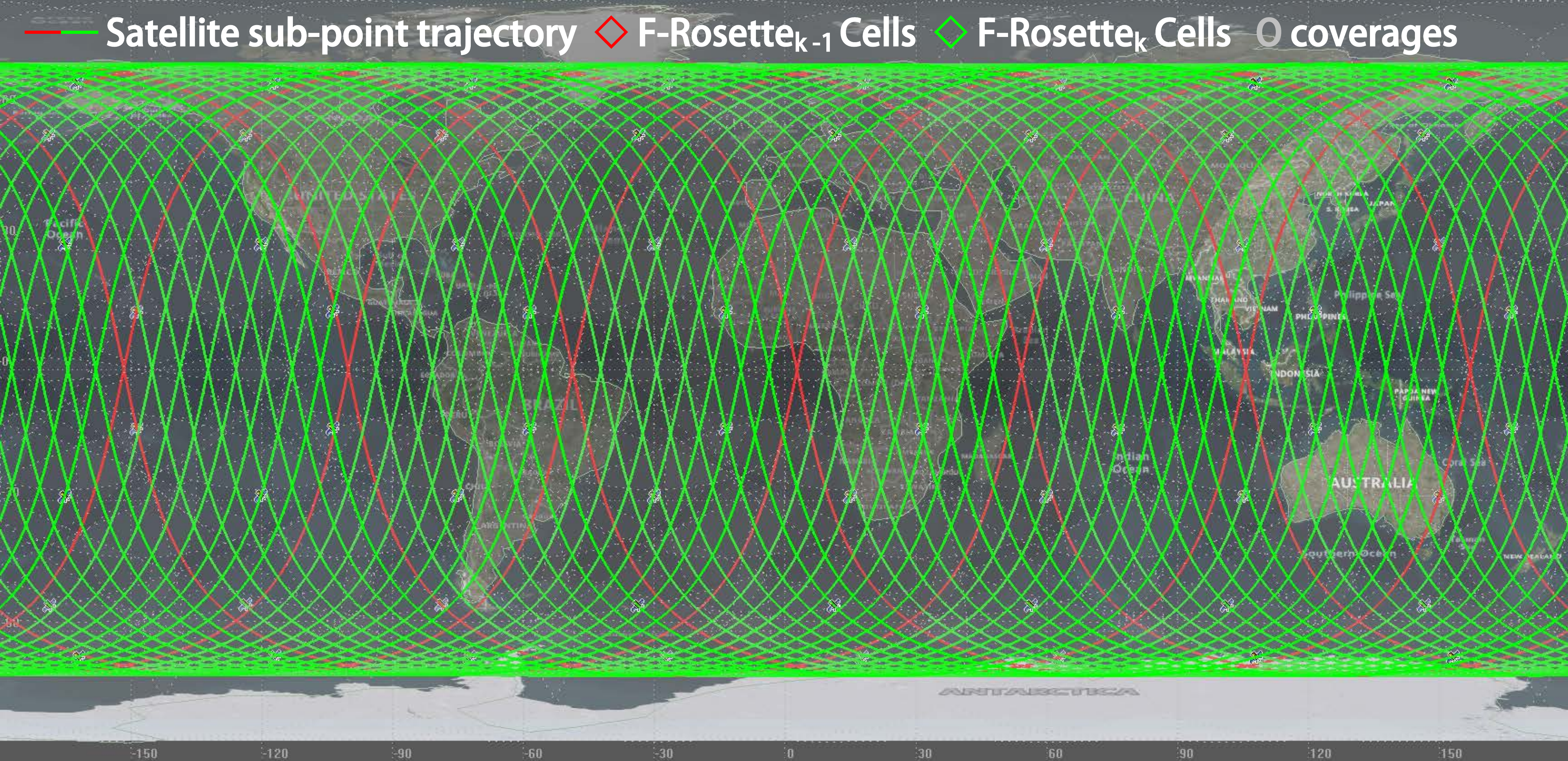}
\label{fig:construction:cell}
}
\vspace{-2mm}
\caption{Recursive construction of \name.}
\label{fig:construction}
\end{figure*}

\begin{algorithm}[t]
   \centering
        \begin{algorithmic}[1]
    	\Require{A Rosette constellation $(N,N,m)$, and a \name level $k\geq0$}
	\Ensure{$\name_k$}
	\State $\name_0=(N,N,m)$ Rosette constellation;
	\For{j=1 to k}
	\State $\name_j=\{\}$; 
	\For{i= 0 to N-1}  \textcolor{gray}{\Comment{{\em\tiny Step 1: Construct $N$ $\name_{j-1}$s}}}
	\State $\name_{j-1}^i=\textsf{Shift}(\name_{j-1}, \frac{2\pi}{N^j})$;
	\State $\name_j$.append($\name_{j-1}^i$);
	\EndFor
	\For{i=0 to N-1}  \textcolor{gray}{\Comment{{\em\tiny Step 2: Inter-connect $\name_{j-1}$s}}}
	\For{each satellite $S_{i}$ in $\name_{j-1}^i$}
	\State Connect $S_i$ to $S_{(i+1\mod N)}$ in $\name_{j-1}^{(i+1\mod N)}$;
	\EndFor
	\EndFor
	\EndFor
	
	\State \Return $\name_k$;

    \end{algorithmic}
    \caption{\name construction. 
    }
    \label{algo:construction}
\end{algorithm}

\name adopts the Rosette constellation as a base case, but extends it recursively with inter-satellite links. 
Specifically, a $\name_0$ is a $(N,m)$ Rosette constellation in $\S$\ref{sec:rosette-primer}, with inter-satellite links in adjacent satellites (exemplified in Figure~\ref{fig:base-case}) and ground-track repeat orbits for all satellites.
A $\name_k$ ($k\geq1$) is constructed recursively from $N$ $\name_{k-1}$s.
Algorithm~\ref{algo:construction} shows \name construction.
It takes two phases (exemplified in Figure~\ref{fig:construction}):
\yuanjie{The \name construction in Algorithm~\ref{algo:construction} does not mandate the base case in Figure~\ref{fig:base-case}, which only uses local links and results in detouring when inter-orbit links cannot be established (Theorem~\ref{thm:hop}). By extending the base case to non-local links, \name can achieve similar lower latency to motif.}

\paragraphb{(1) Construct $N$ $\name_{k-1}$s:} 
The $\name_{k-1}$s are numbered from 0 to $N-1$ as $\name_{k-1}^0$, $\name_{k-1}^1$, $...$, $\name_{k-1}^{N-1}$. 
As shown in Figure~\ref{fig:shift}, $\name_{k-1}^i$ is obtained by first duplicating a $\name_{k-1}^0$, and then shifting it by a time interval $t_k=\frac{T_S}{N^k}$ (or equivalently a phase angle shift $x_k=\frac{2\pi}{T_s}\cdot t_k=\frac{2\pi}{N^k}$ according to Equation \ref{eqn:x}). 
Each satellite $S_j$ in $\name_{k-1}^0$ is replicated $N$ times on the same orbit, denoted as $S_j^0,S_j^1,...S_j^{N-1}$.

\paragraphb{(2) Interconnect $\name_{k-1}$s:} The next step is to bridge $\name_{k-1}$s with {\em intra-orbit} inter-satellite links only.
As shown in Figure~\ref{fig:logic-topology} and Figure~\ref{fig:3d-constellation}, each $S_i$ in $\name_{k-1}^i$ ($i=0,1,...,N-1$) adds a link to $S_{(i+1\mod N)}$ in $\name_{k-1}^{(i+1\mod N)}$. 
This creates a circle among satellites on the same orbit.









\subsection{Basic Properties of \name}
\label{sec:property}



\name inherits all good properties from the classical Rosette constellation in $\S$\ref{sec:rosette-primer}, and has additional appealing features for networking due to its recursive architecture. 

\paragraphb{Network structure size:}
A $\name_k$ has $N^{k+1}$ satellites and $(k+1)N^{k+1}$ inter-satellite links.
Table~\ref{tab:altitude} exemplifies the scale when $N=8$.
Each satellite has $2(k+1)$ inter-satellite links, which is affordable even for small LEO satellites (typically 4--5 links \cite{handley2018delay,bhattacherjee2019network}). 

\paragraphb{Full coverage:} Each $\name_k$ is also a Rosette constellation and thus can ensure full coverage.
Specifically, we have the following theorem (proved in Appendix~\ref{proof:coverage}):

\begin{thm}[Full coverage]
A $\name_k$ guarantees full coverage if its satellites' altitude $H>H_{min}$, where
\begin{align*}
H_{min}&=R_E\left(\frac{1}{\cos R-\sin R\tan\phi}-1\right)\\
R&=\sec^{-1}\left(\sqrt{3}\tan{\left(\frac{\pi}{6}\frac{N^{k+1}}{N^{k+1}-2}\right)}\right)
\end{align*}
\label{thm:coverage}
\end{thm}
With this full coverage guarantee, a satellite operator can customize the \name size to meet its demands.
In Appendix \ref{on-demand-size}, we show how to select the size given the demand of ground-to-space RTT and the minimum elevation angle. 

\paragraphb{Stable network topology:} 
\name guarantees time-invariant network topology despite satellites' high mobility and earth's rotation.
This is desirable to simplify the routing and ensure high network usability ($\S$\ref{sec:challenge}). 
\name guarantees the logical connection between any two satellites remains always-on and unchanged.
This can be achieved by specifying the minimum altitude of all satellites, as shown below (proved in Appendix~\ref{proof:stability}):

\begin{thm}[Stable network topology]
The network topology of a $\name_k$ remains stable if its satellites' altitude $H>\max\{(\frac{1}{cos\frac{r_{max}}{2}}-1)R_E,H_{min}\}$, where $r_{max}$ is the max great circle range of inter-orbit links and $R_E$ is the earth radius. 
\label{thm:stability}
\end{thm}

\paragraphb{Stable satellite sub-point trajectory:} \name always adopts the Rosette constellations with ground-track repeat orbits $T=T_E/(N-m)$.
As shown in Figure~\ref{fig:construction:cell}, this results in stable and periodic satellite sub-point trajectory.
As we will see in $\S$\ref{sec:address}, these trajectories enable hierarchical space-ground network addresses and simplify routing. 

\begin{table}[t]
	\caption{\name's satellites' altitude with $N$=8/16, $m$=1, and the elevation angle $\phi=25^{\circ}$.}
	\vspace{-2mm}
	\label{tab:altitude}
	\resizebox{1\columnwidth}{!}{{
			\begin{tabular}{c|c|c|c|c|c}
				\hline
				\multicolumn{2}{c|}{}& {\bf Num.} & {\bf Minimun} & {\bf Ground-to-} & {\bf Avg. cell}\\
				\multicolumn{2}{c|}{}& {\bf satellites} & {\bf altitude (km)} & {\bf space RTT (ms)} & {\bf size (km$^2$)}\\
				\hline
				\multirow{3}{*}{\rotatebox[origin=c]{90}{$N=8$ }}& $\name_0$  &8 & 11848.46 & 78.99 & 9,060,419 \\
				
				&$\name_1$&64 & 1259.58 & 8.40 & 141,569 \\
				
				&$\name_2$ & 512 &  335.33 & 2.23 & 2,212 \\
				
				\hline
				\multirow{3}{*}{\rotatebox[origin=c]{90}{$N=16$ }}&$\name_0$ &16 & 4268.73 & 28.46 & 1,973,158 \\
				
				&$\name_1$ &256 & 504.83 & 3.36 & 7,707 \\
				
				&$\name_2$ & 4096 &  107.62 & 0.72 & 30 \\
				
				\hline
			\end{tabular}
	}}
\end{table}

\paragraphb{Propagation delay of an inter-satellite link:}
Despite high satellite mobility, \name retains {\em regular and predictable} delay variations for each inter-satellite link. 
For intra-orbit links, \name guarantees {\em fixed} propagation delay between two satellites, due to their time-invariant distances.
For the inter-orbit links, its length (thus propagation delay) unavoidably varies over time.
For any two satellites $i$ and $j$ in different orbits, its inter-satellite link length $L_{ij}$ and propagation delay $\tau_{ij}$:
\begin{align}
\label{eqn:isl-distance}
L_{ij}(t) &=2\left ( H+R_{E} \right )\sin\left ( r_{ij}(t)/2 \right ) \\
\tau_{ij}(t)&=L_{ij}(t)/c \label{eqn:rtt}
\end{align}

where $r_{ij}(t)$ is the {\em range} between $i$ and $j$ on the great circle:
\begin{align}
\begin{split}
\sin^2(r_{ij}(t)/2) &= \cos^4(\beta/2)\sin^2(m+1)(j-i)(\pi/N)\\
		      &+ 2\sin^2(\beta/2)\cos^2(\beta/2)\sin^2m(j-i)(\pi/N)\\
		      &+ \sin^4(\beta/2)\sin^2(m-1)(j-i)(\pi/N)\\
		      &+ 2\sin^2(\beta/2)\cos^2(\beta/2)\sin^2(j-i)(\pi/N)\\
		      &\cdot\cos[\frac{4\pi}{T}t+2m(j+i)(\pi/N)]
\end{split}
\label{eqn:rij}
\end{align}
Note such delay is {\em periodic} and {\em bounded}, thus facilitating delay-sensitive apps. We will evaluate the link delay in $\S$\ref{sec:eval:characteristics}.

\section{Network Addressing in \name}
\label{sec:address}

We next define \name's hierarchical network addressing for space and ground. 
We show how \name decouples its network addressing from mobility for stability, and avoids frequent user IP changes in $\S$\ref{sec:challenge}.


\subsection{Hierarchical Addressing for Satellites} 
\label{sec:address:sat}

\name's recursive structure naturally defines {hierarchical} network addresses for its satellites. 
We assign each satellite $S$ in $\name_k$ with a hierarchical address \textsf{$s_0.s_1.s_2...s_{k}$} ($s_i=0,1,...,N-1, \forall i$).
Each digit $s_i$ represents the corresponding $\name_{i}$ structure that $S$ belongs to. 
The length of this satellite address is $k\log N$ bits (which is usually small as evaluated in $\S$\ref{sec:eval:addr}).
This hierarchical address can be naturally embedded into IPv6's address space, as shown in Figure~\ref{fig:sat-address}.
It is {constant} despite satellite mobility, since it only relies on \name's stable network topology.


\subsection{Hierarchical Addressing for Ground}
\label{sec:address:ue}


To stabilize the addressing in high mobility, \name divides the earth's spherical surface into a hierarchy of disjoint {\em geographical cells} and allocates cell-based addresses to terrestrial users and ground stations. 
In this way, the network addresses are decoupled from the satellite and earth mobility, thus free of frequent changes ($\S$\ref{sec:challenge}). 
They are unchanged unless the users move to a new geographical cell. 
Figure \ref{fig:construction:cell} exemplifies \name's cells.
Each cell is a quadrilateral bounded by four satellite sub-point trajectories of \name's satellites. 
The \name cells exhibit three appealing features for satellite routing in space:


\paragraphb{Stable geographical cells:}
\name's satellites run on the ground-track orbits ($\S$\ref{sec:construction}), which implies all satellite sub-point trajectories are {\em time-invariant}.
Therefore, despite the high satellite mobility and earth's rotation, each \name cells's location, size, and coverage are {fixed} and invariant. 

\begin{figure}[t]
\vspace{-15mm}
\subfloat[Base case (k=0)]{
\includegraphics[width=.7\columnwidth]{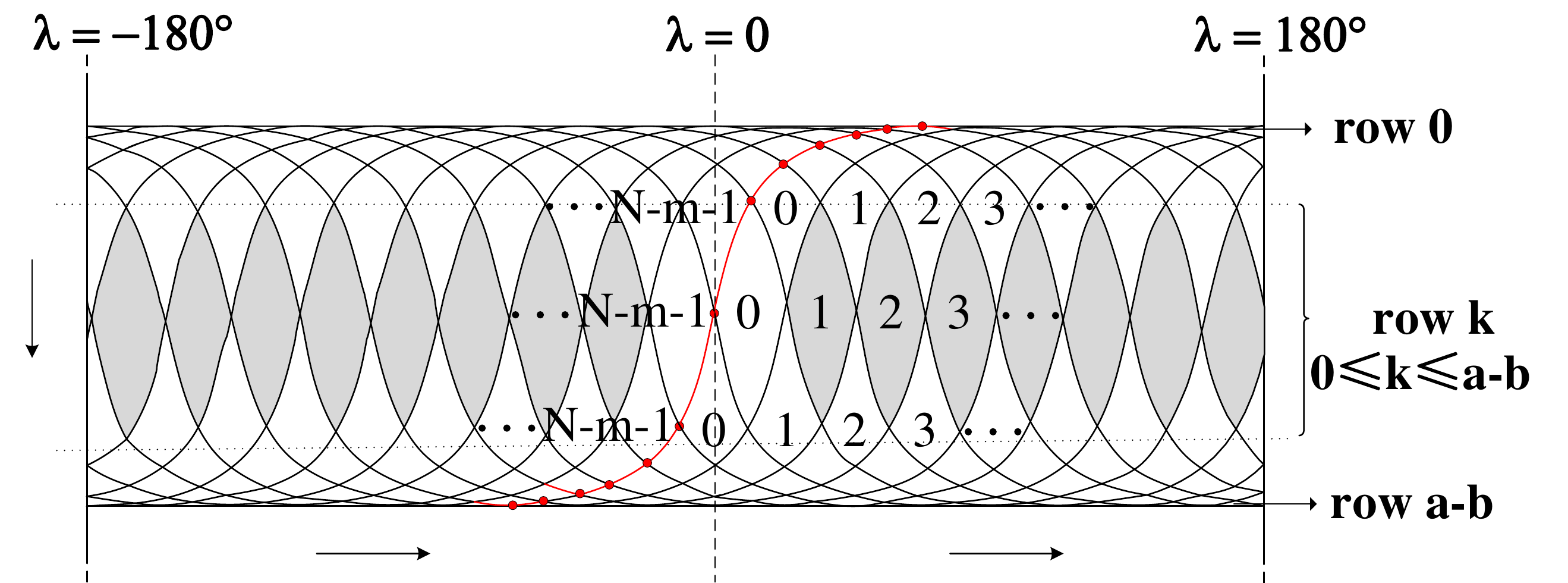}
\label{fig:big-cell}
}
\subfloat[Hierarchical cells]{
\includegraphics[width=.3\columnwidth]{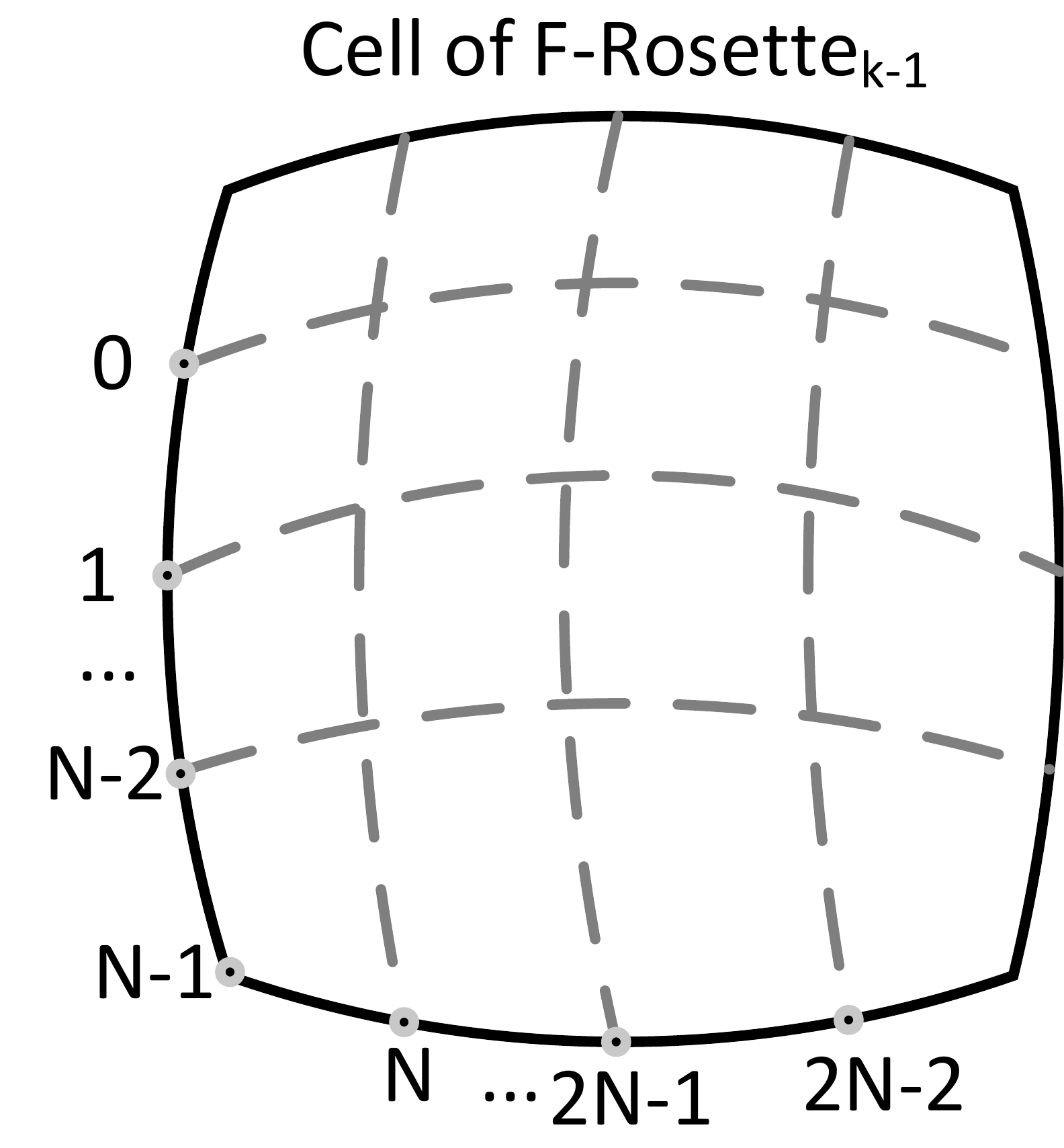}
\label{fig:sub-cell}
\label{fig:proof-hierarchy}
}
\caption{Hierarchical geographical cells in \name.}
\end{figure}

\begin{figure}[t]
\centering
\vspace{-5mm}
\subfloat[Satellite address]{
\includegraphics[width=\columnwidth]{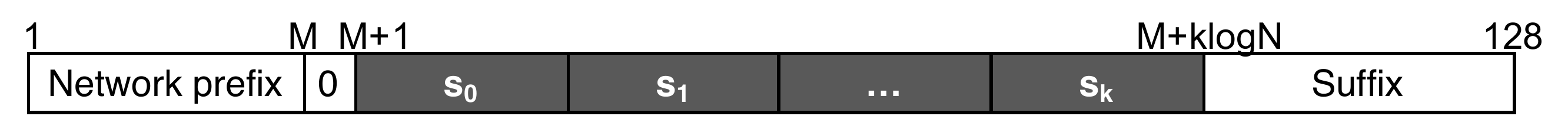}
\label{fig:sat-address}
}
\vspace{-4mm}
\subfloat[Ground address]{
\includegraphics[width=\columnwidth]{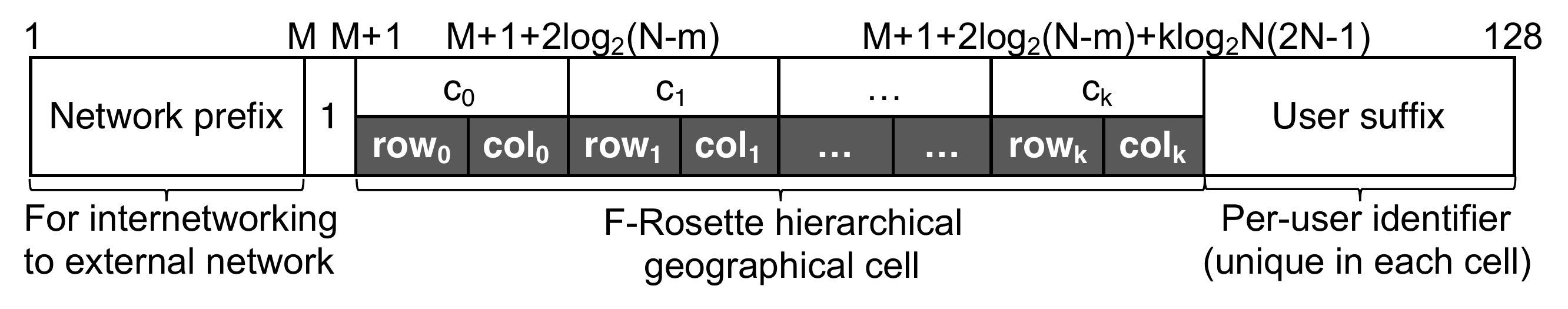}
\label{fig:user-address}
}
\caption{Embedding \name into IPv6 address. 
\lixin{Add fields or opt-headers to optimize routing.}
}
\label{fig:address}
\end{figure}

\paragraphb{Hierarchical geographical cells:}
\name's recursive and symmetric structure forms a hierarchy of cells.  
With recursive construction, each cell in the $\name_{k-1}$ will be divided into a group of sub-cells in $\name_{k}$.
This hierarchy is exemplified in Figure~\ref{fig:proof-hierarchy} and validated by the following theorem (proved in Appendix~\ref{proof:hierarchy}): 

\begin{thm}[Hierarchical geographical cells]
When constructing $\name_k$ from $\name_{k-1}$, each geographical cell in $\name_{k-1}$ will be divided in to $N^2$ disjoint sub-cells in $\name_k$.
\label{thm:hierarchy}
\end{thm}

We also have the total number of cells in a $\name_k$ as follows (proved in Appendix~\ref{proof:cells}): 
\begin{lem}
A $\name_k$ divides its coverage areas into $(N-m)^2N^{2k}$ disjoint cells. 
\label{lem:cells}
\end{lem}

\paragraphb{Network addressing for the ground:}
\name assigns geographical address to each ground user based on the cell it resides. 
Figure~\ref{fig:user-address} illustrates how this geographical cell ID can be readily realized embedded into IPv6\footnote{Besides IPv6, this address can also be embedded into the future architecture such as NDN~\cite{zhang2014named,jacobson2009networking}, SCION \cite{zhang2011scion}, MobilityFirst \cite{mobilityfirst}, and XIA \cite{han2012xia}.}.
For each user in this cell, \name assigns its address as a concatenation of new prefix, hierarchical geographical ID, and the user suffix.
The network prefix is used for inter-networking with external networks and backward compatibility with standard IPv6 ($\S$\ref{sec:deployment}).
The per-user suffix guarantees the globally unique address inside each geographical cell. 
In this way, the user address remains {\em invariant} unless (s)he moves to a new cell (which is rare due to the cell size in Table~\ref{tab:altitude}).
To help detect if the address is for satellite or ground for routing in $\S$\ref{sec:routing}, we add a 1-bit flag to differentiate their addresses.

We next elaborate on how to allocate the hierarchical geographical cell IDs.
A geographical cell in $\name_k$ is uniquely identified by $c_0.c_1...,.c_{k}$, where $c_0=0,1,...(N-m)^2-1$   
and $c_i=0,1,...,N^2-1 (i\neq0)$. 
Each digit $c_i$ represents the corresponding $\name_{i}$ cell identifier that this user belongs to, and can be further decomposed into a row ID $row_i$ and column ID $col_i$. 
For $\name_0$, its sub-satellite point trajectories and two latitude lines divide the earth into $N-m$ row and $N-m$ columns, and we identify each cell with the left point. 
As shown in Figure~\ref{fig:big-cell}, we number each row from $0$ to $N-m-1$ from north to south. 
Starting from the cell whose left dot is on the trajectory of satellite 0 above the equator and its extension line, each column is numbered from $0$ to $N-m-1$ according to the increasing direction of the longitude value.
Next, a $\name_k$ ($k\geq1$) divides each cell in $\name_{k-1}$ $\left( k\ge 1 \right) $ into $2N-1$ layers. We number each layer from 0 to $2N-2$ in the direction from north to south, and at the same time, we number cells in each layer from left to right.
Encoding this hierarchical geographical cell address requires $2k\log N+2log[(N-m)]$ bits (which is usually small for IPv6, as evaluated in $\S$\ref{sec:eval:addr}).

\section{Routing in \name}
\label{sec:routing}

We describe how \name enables stable, highly-usable, and efficient routing \add{without re-convergence ($\S$\ref{sec:challenge})}. 
We first show how to route between satellites over \name's stable graphical topology ($\S$\ref{sec:routing-sat}).
Then we show how to route between ground users by embedding hierarchical geographical routing into the satellites' topological routing ($\S$\ref{sec:routing-user}). 


\subsection{Topological Routing for the Satellites}
\label{sec:routing-sat}

With its stable and regular constellation, \name supports efficient and highly-usable single/multi-path routing that is compatible with standard Internet routing mechanisms. 

\begin{algorithm}[t]
   \centering
        \begin{algorithmic}[1]
    	\Require{Source satellite address $S$=\textsf{$s_0.s_1.s_2...s_{k}$}, destination satellite address $D$=\textsf{$d_0.d_1.d_2...d_{k}$}, and a permutation $P=[p_0,p_1,...,p_{k}]$ of [0,1,...,k]}
	\Ensure{A path from $S$ to $D$}	
	\State \textsf{path(S,D)=\{S,\}};  \textsf{next\_hop=S};
	\For{i=$p_0,p_1,...,p_{k}$} \textcolor{gray}{\Comment{{\em\tiny The permutation shuffles indexes for multi-path routing}}}
		\If{\textsf{next\_hop[i] $-d_i$ mod N$\leq N/2$}} \textcolor{gray}{\Comment{{\em\tiny Clockwise routing is shorter}}}
		\While{\textsf{next\_hop[i] $\neq d_i$}}
			\State \textsf{next\_hop[i] = next\_hop[i]+1 mod N}; 
			\State \textsf{path(S,D).append(next\_hop)};
		\EndWhile
		\Else \textcolor{gray}{\Comment{{\em\tiny Counter-clockwise routing is shorter}}}
		\While{\textsf{next\_hop[i] $\neq d_i$}}
			\State \textsf{next\_hop[i] = next\_hop[i]-1 mod N};
			\State \textsf{path(S,D).append(next\_hop)};
		\EndWhile
		\EndIf
	\EndFor
	\State \Return \textsf{path};
    \end{algorithmic}
    \caption{Shortest-path satellite routing in \name.}
    \label{algo:single-path-routing}
\end{algorithm}

\begin{figure}[t]
\vspace{-6mm}
\includegraphics[width=.9\columnwidth]{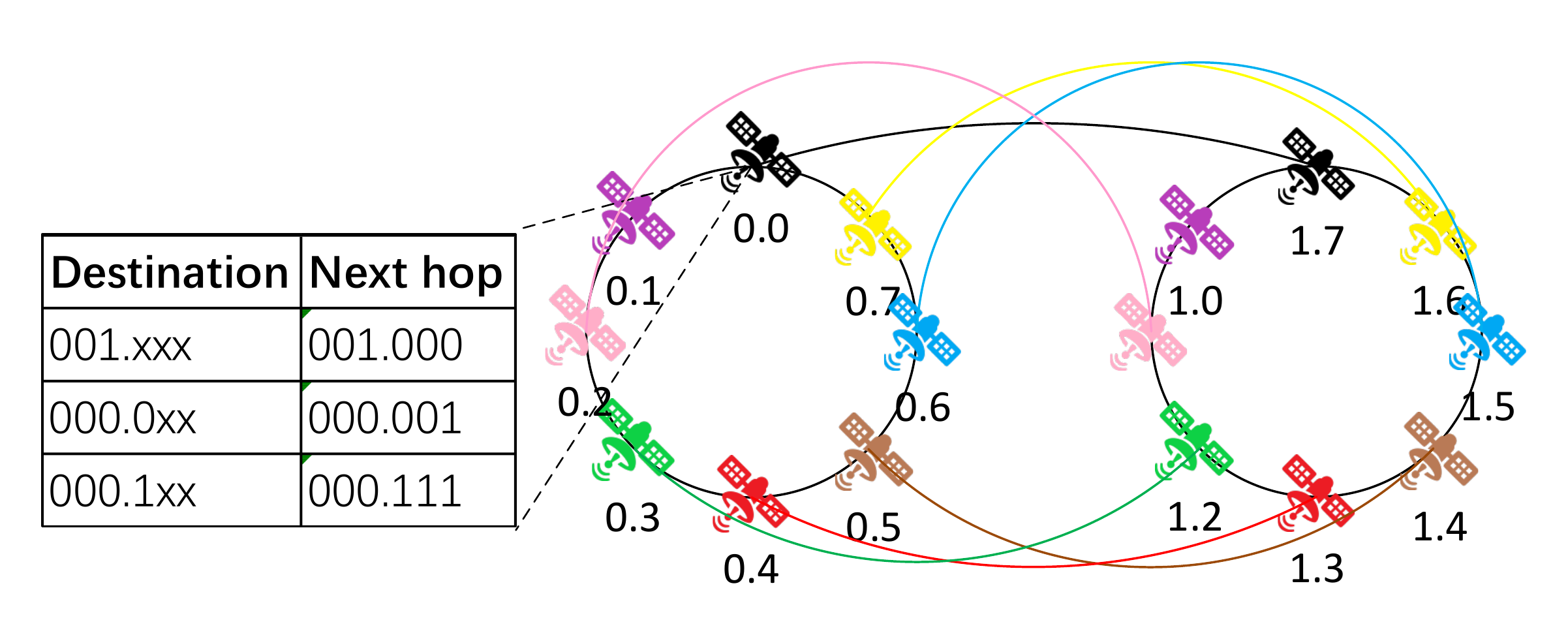}
\caption{Prefix/wildcard match in \name routing.}
\label{fig:fib}
\end{figure}

\paragraphb{Single-path routing:}
With \name's regular topology, a satellite can {\em locally} compute the shortest path without global routing (re-)convergence and thus avoids low network usability in $\S$\ref{sec:challenge}.
Algorithm~\ref{algo:single-path-routing} shows the shortest-path routing from a source satellite $S$=\textsf{$s_0.s_1.s_2...s_{k}$} to a destination $D$=\textsf{$d_0.d_1.d_2...d_{k}$} defined in $\S$\ref{sec:address:sat}.
A pre-defined permutation $P=[p_0,p_1,...,p_{k}]$ of indexes 0,1,...,k is also given as an input to facilitate the enumeration of multiple paths (detailed below). 
To route traffic from $S$ to $D$, we follow the permutation $P$ to traverse all layers of $\name_k$.
At each layer $p_i$, all satellites form a ring topology according to $\S$\ref{sec:construction}.
The shortest path is either the clockwise or counter-clockwise routing.
To find the shortest path, we first decide the routing direction, and then increment (or decrement) the $p_i$-th digit of the previous satellite's address, until the $p_i$-th digit equals the destination's corresponding digit $d_{p_i}$.
The correctness of Algorithm~\ref{algo:single-path-routing} is therefore straightforward: 

\begin{thm}[Shortest path]
\label{thm:shortest-path}
Algorithm~\ref{algo:single-path-routing} finds the shortest path in hop counts between any two satellites.
\end{thm}

\name's routing exhibits three desirable features: 

\paragraphe{(1) Stable and highly usable without re-convergence:}
Algorithm~\ref{algo:single-path-routing} avoids global routing (re-)convergence in \name's regular and stable topology, thus retain high usability despite dynamic many-to-many mapping ($\S$\ref{sec:challenge}).
With \name's regular structure, the satellite's address inherently implies the routing path.
Each satellite can locally decide the next hop without global routing (re-)convergence.

\paragraphe{(2) Prefix/wildcard-based routing table:}
With \name's hierarchical satellite address, Algorithm~\ref{algo:single-path-routing} can be realized with the standard prefix/wildcard matching in IP networks.
Figure~\ref{fig:fib} exemplifies the routing table for Algorithm~\ref{algo:single-path-routing}.
With the ring topology at each \name layer, destinations with the same prefix (or layer) can be aggregated as a single entry in the standard routing table. 
The storage required by the \name routing table is also small for memory-constrained satellites (proved in Appendix~\ref{proof:fib}):

\begin{thm}[FIB storage bound]
Each satellite in a $\name_k$ has at most $2(k+1)\lceil\log\frac{N}{2}\rceil$ entries in its routing table. 
\label{thm:fib}
\end{thm}

\paragraphe{(3) Performance bound:}
\name's regular structure also provides bounded latency for its single-path routing.
The upper bound can be derived from Algorithm~\ref{algo:single-path-routing} as follows:

\begin{thm}[Hop bound]
\label{thm:hop}
The maximum number of hops between two satellites in $\name_k$ is no more than $\frac{(k+1)N}{2}$. 
\end{thm}

\paragraphb{Multi-paths:}
\name also offers abundant parallel paths for performance acceleration and fault tolerance.
By enumerating the permutations $P$ of the \name layers [0,1,...,k], Algorithm~\ref{algo:single-path-routing} yields parallel paths from the same source-destination pair.
The following theorem shows the available paths in \name (proved in Appendix~\ref{proof:multipath}):
\yuanjie{These multi-paths propogation delays are still different, and vary regularly.}

\begin{thm}[Multi-path]
There are $2(k+1)$ disjoint parallel paths between any two satellites in a $\name_k$.
\label{thm:multipath}
\end{thm}

\paragraphb{Stability and high network usability:}
With its stable network topology, \name ensures stable routing despite high mobility.
Each satellite locally routes traffic without requiring slow global (re)-convergence. 
This prevents the low network usability as discussed in $\S$\ref{sec:challenge}.

\subsection{Geographical Routing for the Ground}
\label{sec:routing-user}

\name adopts geographical routing to deliver traffic between terrestrial users.
Geographical routing is appealing for its efficiency, scalability, and high network usability.
It takes the physically short path to forward the traffic. 
Moreover, geographical routing can be mostly performed locally without global routing (re-)convergence, thus avoiding low network usability in mobility ($\S$\ref{sec:challenge}).

\begin{figure}[t]
\vspace{-10mm}
\includegraphics[width=\columnwidth]{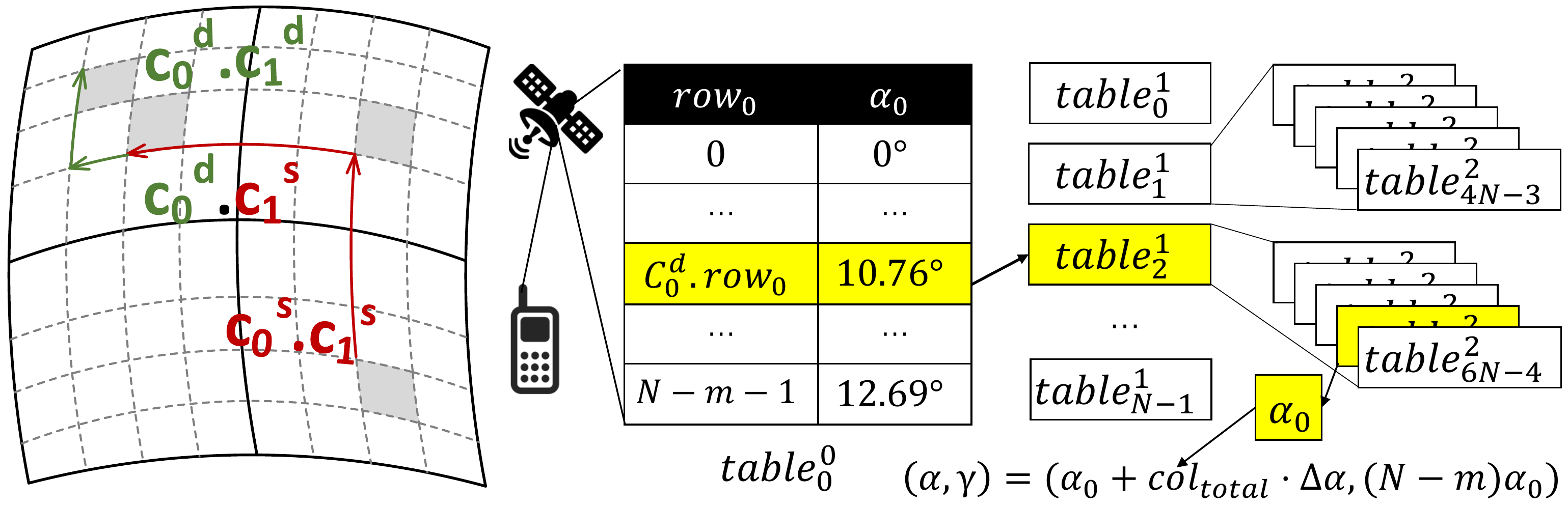}
\vspace{-2mm}
\caption{Hierarchical geographical routing for users.}
\label{fig:geo-routing}
\end{figure}

\name supports native {\em hierarchical} geographical routing for the ground.
As shown in Figure~\ref{fig:geo-routing}, data from cell $c_0^s.c_1^s....c_{k}^s$ to $c_0^d.c_1^d....c_{k}^d$ in $\S$\ref{sec:address:ue} can traverse from $c_i^s$ to $c_i^d$ at each level $i$ of the \name cells.
In \name, this geographical routing can be converted to an equivalent topological routing between satellites in $\S$\ref{sec:routing-sat}, thus retaining stability and efficiency.
We next elaborate on it. 

\begin{algorithm}[t]

   \centering
        \begin{algorithmic}[1]
        \Require{Source user's serving satellite's $S=(\alpha_s(t),\gamma_s(t))$ and destination user's cell $D=(\alpha_d,\gamma_d)$}
	\Ensure{A routing path from $S$ to $D$ through \name satellites}	
	\State \textsf{path(S,D)=\{S,\}};  \textsf{next\_hop=S};
	\State $\Delta\alpha=\alpha_d-\alpha_s(t)\mod 2\pi$; $\Delta\gamma=\gamma_d-\gamma_s(t)\mod 2\pi$; 
	\If{$\Delta\alpha<\pi$} \textcolor{gray}{\Comment{{\em\tiny Clockwise routing is shorter}}}
		\While{$\Delta\alpha\geq\frac{2\pi}{N}$}
		\State \textsf{next\_hop[0] = next\_hop[0]+1 mod N};
		\State $\Delta\alpha=\Delta\alpha-\frac{2\pi}{N}\mod 2\pi$; $\Delta\gamma=\Delta\gamma-\frac{2m\pi}{N}\mod 2\pi$;
		\State \textsf{path(S,D).append(next\_hop)}; {goto Line~\ref{ln:return} if \textsf{next\_hop} covers $D$;}
		\EndWhile
	\Else \textcolor{gray}{\Comment{{\em\tiny Counter-clockwise routing is shorter}}}
		\While{$\Delta\alpha\geq\frac{2\pi}{N}$}
		\State \textsf{next\_hop[0] = next\_hop[0]-1 mod N};
		\State $\Delta\alpha=\Delta\alpha+\frac{2\pi}{N}\mod 2\pi$; $\Delta\gamma=\Delta\gamma+\frac{2m\pi}{N}\mod 2\pi$;
		\State \textsf{path(S,D).append(next\_hop)}; {goto Line~\ref{ln:return} if \textsf{next\_hop} covers $D$;}
		\EndWhile
	\EndIf
	
	\For{i=1,2,...,k-1}
	\If{$\Delta\gamma<\pi$} \textcolor{gray}{\Comment{{\em\tiny Clockwise routing is shorter}}}
		\While{$\Delta\gamma\geq\frac{2m\pi}{N^k}$}
		\State \textsf{next\_hop[i] = next\_hop[i]+1 mod N};
		\State $\Delta\gamma=\Delta\gamma-\frac{2m\pi}{N^{k}}\mod 2\pi$;
		\State \textsf{path(S,D).append(next\_hop)}; {goto Line~\ref{ln:return} if \textsf{next\_hop} covers $D$;}
		\EndWhile
	\Else \textcolor{gray}{\Comment{{\em\tiny Counter-clockwise routing is shorter}}}
		\While{$\Delta\gamma\geq\frac{2m\pi}{N^k}$}
		\State \textsf{next\_hop[i] = next\_hop[i]-1 mod N};
		\State $\Delta\gamma=\Delta\gamma+\frac{2m\pi}{N^{k}}\mod 2\pi$;
		\State \textsf{path(S,D).append(next\_hop)}; {goto Line~\ref{ln:return} if \textsf{next\_hop} covers $D$;}
		\EndWhile
	\EndIf

	\EndFor
	\State \Return \textsf{path}; \label{ln:return}	
    \end{algorithmic}
    
    \caption{\name's geographical routing for ground.}
    \label{algo:geo-routing}
\end{algorithm}

\paragraphb{{Space-ground alignment with \name coordinates}:}
The first step for \name's geographical routing is to align the logical satellite address in the cyberspace and ground locations in the physical world.
\name's structure implies a geographic coordinate system that facilitates this. 
Figure~\ref{fig:rosette-constellation} exemplifies this coordinate system.
Different from the classical Cartesian coordinates or latitudes/longitudes, 
\name defines the position of each location on the earth as $(\alpha, \gamma)$,
where $\alpha$ is its right ascension angle, and $\gamma$ is the phase angle along a {virtual} orbit of inclination $\beta$ across the right ascension angle.
This coordinate aligns the ground location with \name's satellite sub-point trajectories, thus facilitating the geographical-to-topological routing below.
In this coordinate, each satellite's sub-point location $(\alpha, \gamma)$ can be mapped to the runtime latitude $\varphi$ and longitude $\lambda$ as follows (derived from Equation \ref{eqn:alpha}, \ref{eqn:gamma}, \ref{eqn:phi}, and \ref{eqn:lambda}): 
\begin{align*}
\sin \varphi =\sin \beta \sin \left( N\omega _Et-\gamma \right)  \\
\tan \left( \lambda +\omega _Et-\alpha \right) =\cos \beta \tan \left( N\omega _Et-\gamma \right) 
\end{align*}
Due to the high mobility, a satellite sub-point's longitude shifts over time with dynamic mapping between $(\alpha,\gamma)$ and $(\varphi,\lambda)$.
Despite so, it is easy to verify from Equation \ref{eqn:alpha}--\ref{eqn:x} that, the {\em relative position} $(\Delta\alpha, \Delta\gamma)$ between any two connected satellites' sub-points remains {\em constant}.
This results in an important property to bridge the topological routing between satellites and geographical routing between ground cells:

\begin{prop}[Space-ground routing alignment]
A packet routed through an inter-orbit satellite link in \name moves a fixed relative position of
$$\Delta\alpha_0=\frac{2\pi}{N}, \Delta\gamma_0=\frac{2\pi m}{N}$$
A packet routed through an intra-orbit satellite link at layer $j=1,2,...k$ moves a fixed relative position of
$$\Delta\alpha_j=0, \Delta\gamma_j=\frac{2\pi m}{N^{k}}$$
\label{prop:routing}
\end{prop}

\paragraphb{Geographical-to-topological routing embedding:}
Property \ref{prop:routing} implies routing between satellites in Algorithm~\ref{algo:single-path-routing} is equivalent to traversing a {\em fixed} relative position in \name's coordinate system. 
Given two cells $S=c_0^s.c_1^s....c_{k}^s$ and $D=c_0^d.c_1^d....c_{k}^d$, one can compute their relative position $(\Delta\alpha, \Delta\gamma)$ via table lookup (detailed below), and route the traffic from $S$ to $D$ by following the topological routing in Algorithm~\ref{algo:single-path-routing} to hierarchically shorten the packet's relative position to $D$.
This is equivalent to the geographical routing in Figure~\ref{fig:geo-routing}.

Algorithm~\ref{algo:geo-routing} elaborates on the geographical ground routing via topological satellite routing.
It extends Algorithm~\ref{algo:single-path-routing}, and maps the relative motion to hop count. 
It uses the source user's serving satellite's sub-point as the input, which is locally maintained by the satellite as $\alpha_s(t)=\alpha_s(0)-\frac{2\pi}{T_E}t,\gamma_s(t)=\gamma_s(0)+\frac{2\pi}{T}t$ (Equation~\ref{eqn:alpha}--\ref{eqn:x}).
The routing ends if the user on in target cell is covered by the current satellite (detected if the current satellite finds the user associates to it at the link layer). 
Algorithm~\ref{algo:geo-routing} can be realized atop Algorithm~\ref{algo:single-path-routing}, thus retaining high efficiency for resource-constrained satellites.

\paragraphb{Avoiding local minimum and last-hop ambiguity:}
A classical problem for the greedy geographical routing is the local minimum \cite{ruehrup2009theory, karp2000gpsr, kim2005geographic, leong2006geographic}, \ie, the physically nearest satellite is not reachable for delivery (mostly due to limited coverage). 
This problem is largely mitigated in \name: Similar to greedy embedding \cite{kleinberg2007geographic, lam2012geographic}, geographical routing in Algorithm~\ref{algo:geo-routing} relies on topological routing in $\S$\ref{sec:routing-sat} for guaranteed reachability and successful delivery. 
The only exception is the last hop, where the terrestrial user does not associate to the nearest satellite. 
Since the ground-satellite link depends on user-specific satellite selection strategy and is beyond the satellite topology, simply relying on topological routing is insufficient to guarantee delivery.
Fortunately, Algorithm~\ref{algo:geo-routing} over \name's ring structure implies all satellites at each layer can be reached in the worst case. 
This ensures eventually the user-selected satellite will be reached for successful delivery, at some cost of detouring delays.   

\paragraphb{Data forwarding acceleration via table lookup:}
Similar to the topological routing,
geographical routing in Algorithm~\ref{algo:geo-routing} can also be accelerated via table lookup at least four aspects.
First, the relative motion between connected satellites $(\Delta\alpha_k,\Delta\gamma_k)$ can be pre-computed and pre-stored in each FIB entry, thus avoiding runtime per-packet computation in Algorithm~\ref{algo:geo-routing}.
Second, to decide the destination cell's location $(\alpha_d,\gamma_d)$ given its cell ID $c_0^d.c_1^d.....c_k^d$, 
each satellite can pre-compute and store a hierarchical lookup table as shown in Figure~\ref{fig:geo-routing}. 
Since all geographical cells are formed by the satellite sub-points, one can follow Equation~\ref{eqn:phi}--\ref{eqn:lambda} to pre-compute each cell's location for hierarchical lookup.
This results in a table with $\frac{\left( N-m \right) N^k}{2}$ entries and $O(k+1)$ runtime lookup time (elaborated in Appendix~\ref{id-to-location-table}).
Third, to avoid repetitively computing the source user's serving satellite's runtime location $(\alpha_s(t),\gamma_s(t))$ for all packets,
the satellite can cache its runtime location and refresh it every $\frac{T}{2N^k}$ seconds (which guarantees consistent forwarding decision given the minimal $(\Delta\alpha_k,\Delta\gamma_k)$). 
Last but not least, to avoid repetitive next-hop calculation for the packets to the same destination cell, a satellite can create a cached geographical FIB entry and periodically update it every $\frac{T}{2N^k}$ seconds.

\paragraphb{Stability and high network usability:}
\name embeds geographical routing into \name's stable network topology, thus facilitating stable routing.
Similar to topological routing in $\S$\ref{sec:routing-sat}, \name's geographical routing relies on local information only.
It does not require the slow routing (re)-convergence, thus retaining high network usability despite high satellite mobility and earth's rotation ($\S$\ref{sec:challenge}).

\section{Issues in Practical Deployment}
\label{sec:deployment}

In this section, we discuss various issues and solutions for the deployment of \name in practice.

\yuanjie{Except \ref{sec:deployment:incremental}, other aspects are optional. Decide whether to include them based on space availability.}

\save{
\begin{figure}[t]
\subfloat[Hierarchical expansion]{
\includegraphics[width=0.59\columnwidth]{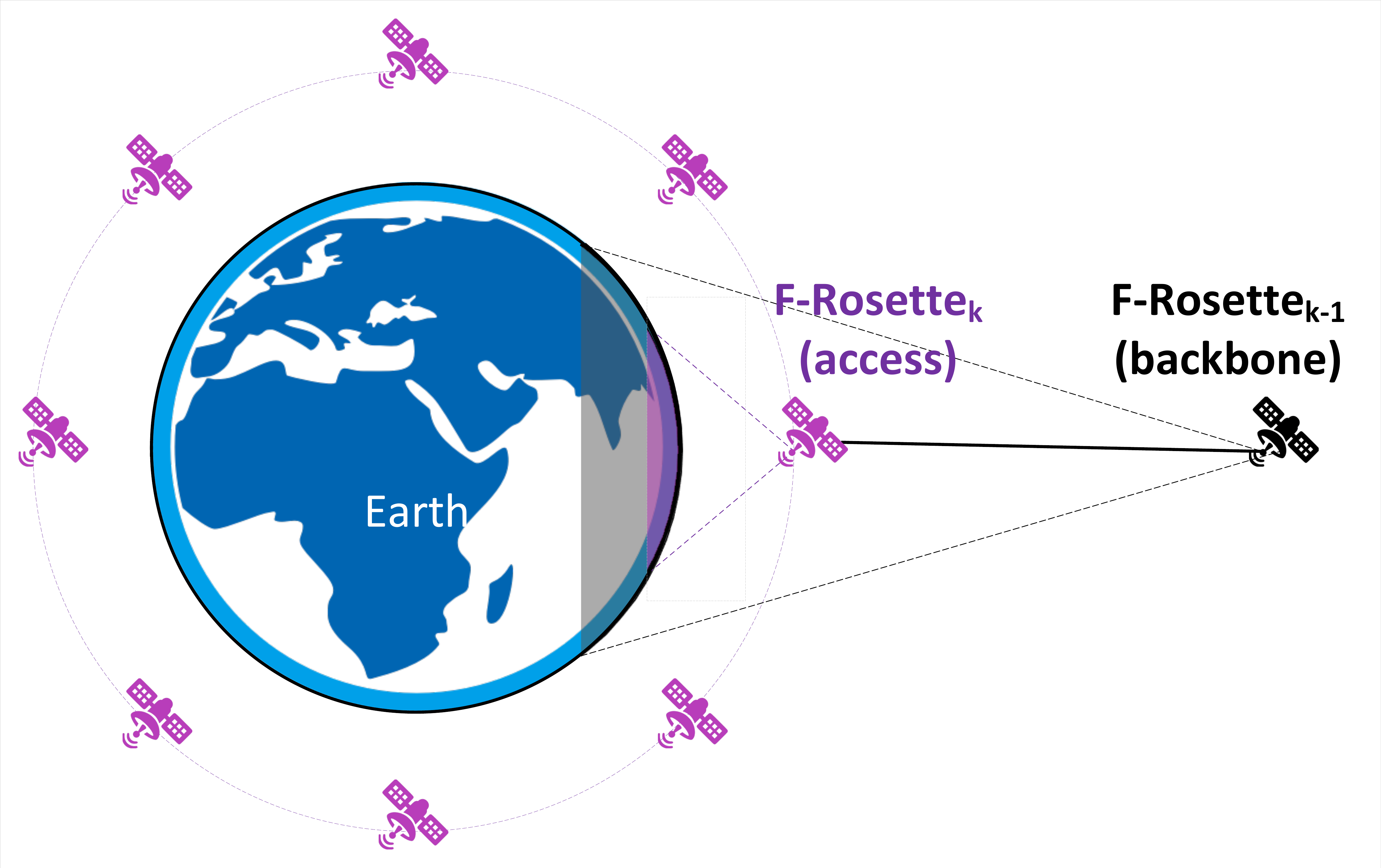}
\label{fig:deployment:hierarchical}
}
\subfloat[Flat expansion]{
\includegraphics[width=0.41\columnwidth]{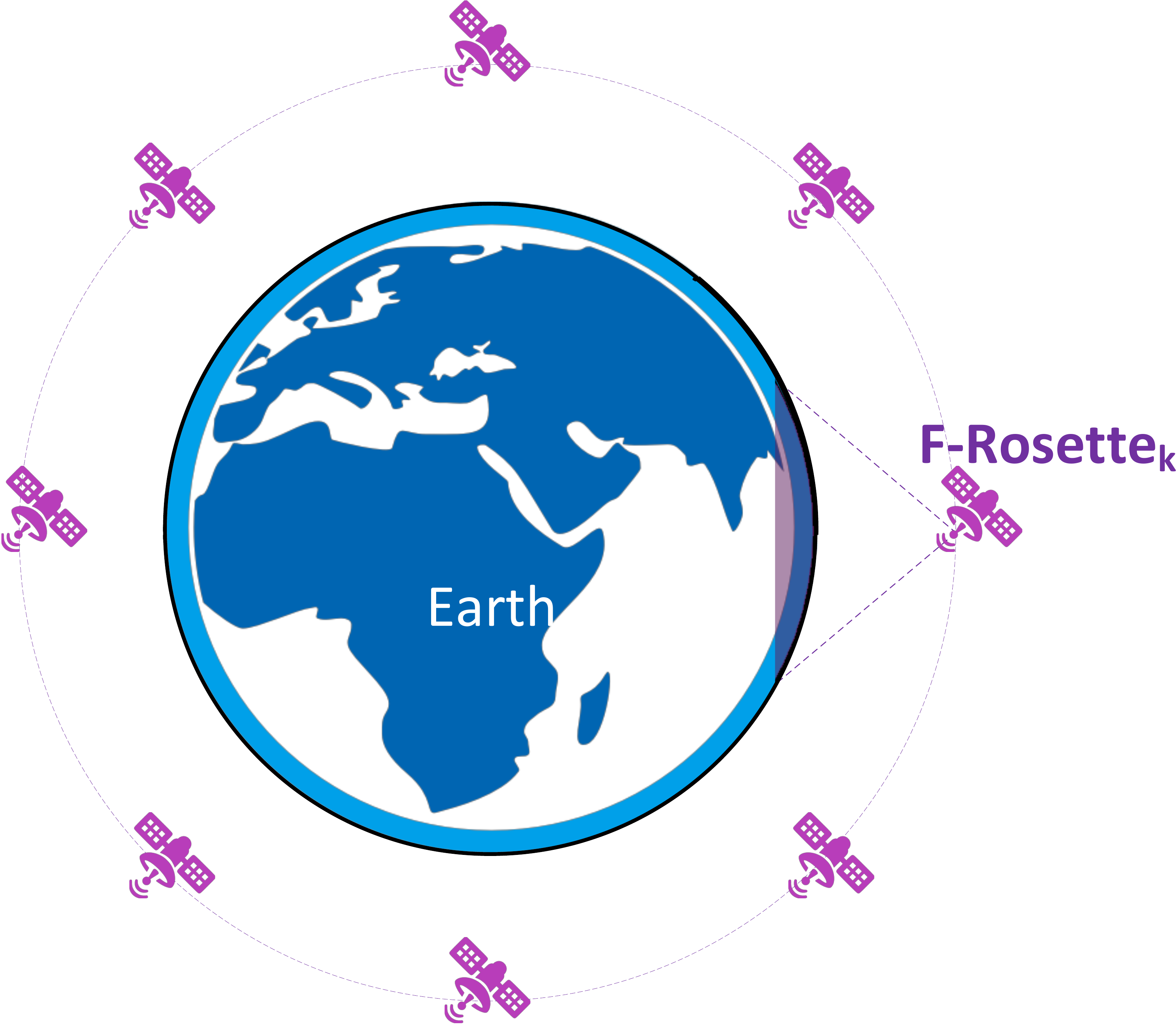}
\label{fig:deployment:flat}
}
\caption{Two incremental expansions in \name.}
\label{fig:deployment}
\end{figure}
}

\begin{figure}[t]
\vspace{-15mm}
\subfloat[$t=0$]{
\includegraphics[width=0.45\columnwidth]{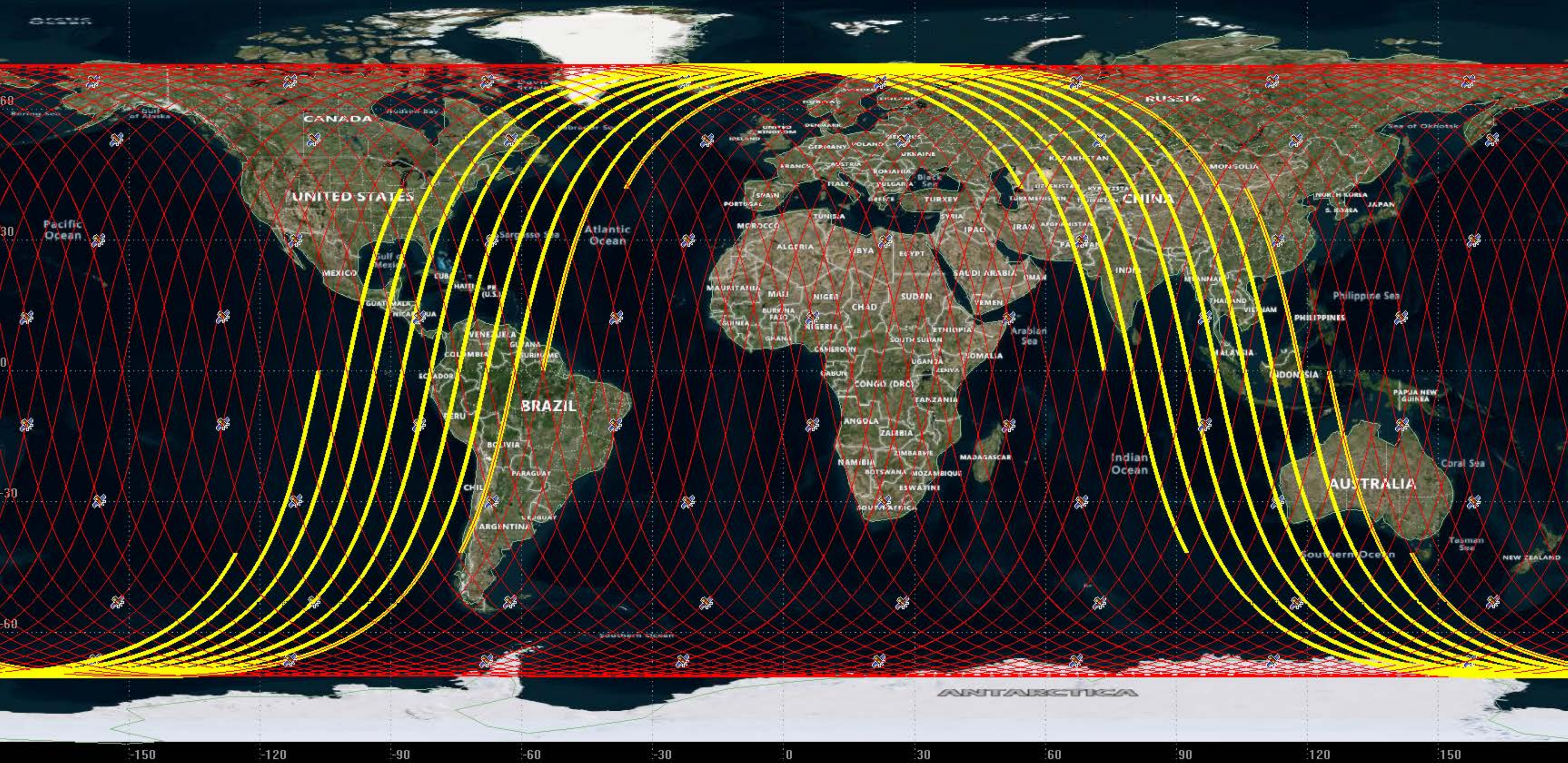}
\label{fig:deployment:t0}
}
\subfloat[$t=T/N$]{
\includegraphics[width=0.45\columnwidth]{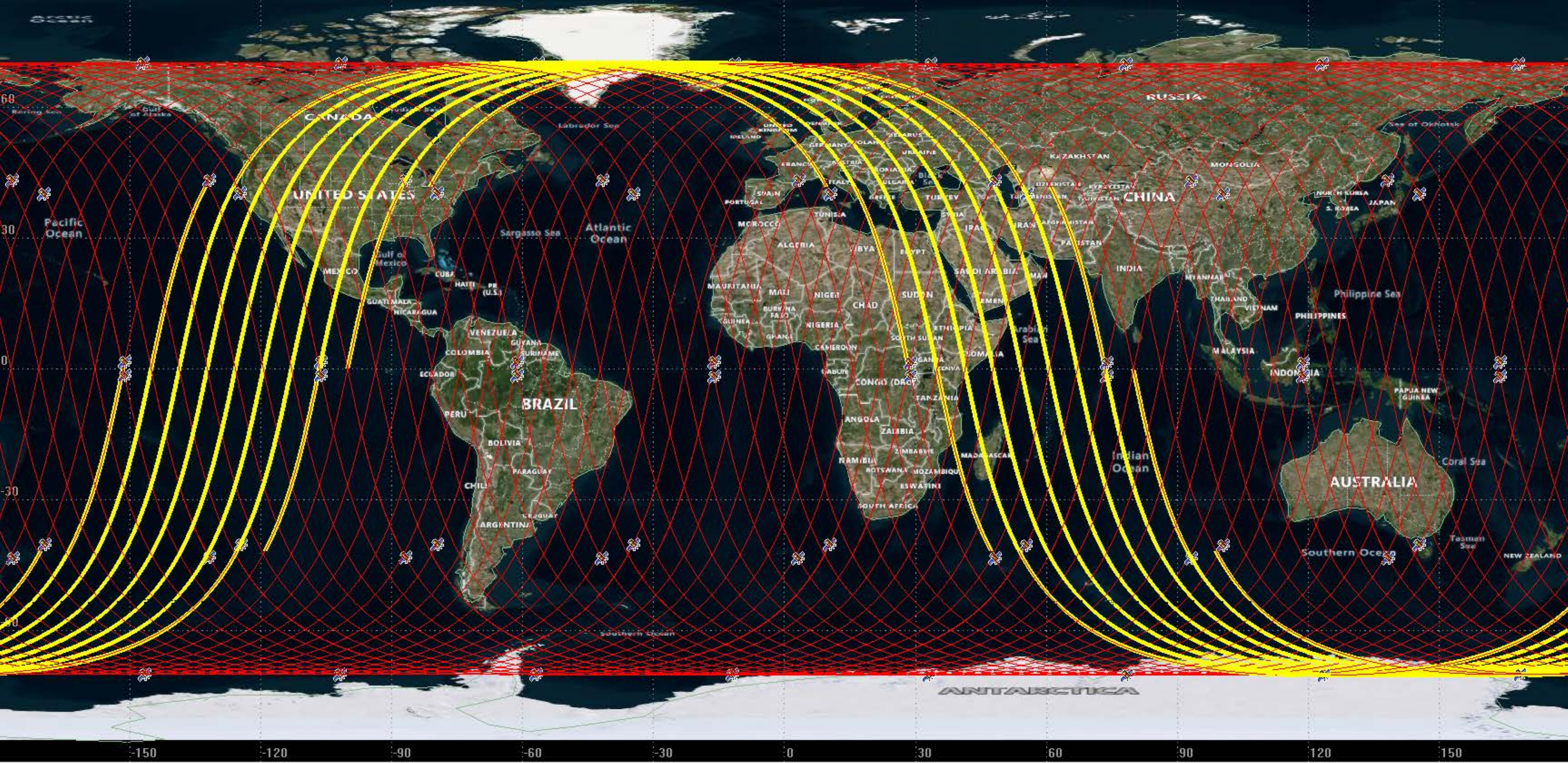}
\label{fig:deployment:t1}
}
\vspace{-2mm}
\caption{Exploiting regular satellite mobility to guarantee data delivery in flat incremental expansion.}
\label{fig:iterative-expansion}
\end{figure}

\paragraphb{Incremental expansion:}
In practice, it is unlikely to deploy \name all at once. The satellites can only be launched to space incrementally. 
For example, as of January 2021, the latest SpaceX Falcon 9 rocket can launch up to 143 satellites each time \cite{rocket-capability}, which is less than $\name_1$'s 256 satellites when $N=16$.
When incrementally expanding \name in space, it is desirable to retain always-on, seamless network services for existing users. 
This demand results in three concrete goals in expansion: (1) guarantee data delivery despite incomplete deployment; (2) avoid re-interconnection of satellites whenever possible; and (3) retain stable address during the deployment. 
\save{To this end, we devise two incremental expansion approaches for \name (exemplified in Figure~\ref{fig:deployment}): A recursive {\em hierarchical deployment}, and an iterative {\em flat deployment}.

\paragraphb{Recursive hierarchical expansion:}
The recursive construction of \name in Algorithm~\ref{algo:construction} implies a natural hierarchical expansion: Deploy $\name_{k-1}$ at higher altitude first, and then use it as a {\em backbone} to connect satellites in $\name_{k}$ at lower altitude. 
When $\name_k$ is being expanded, the high-altitude $\name_{k-1}$ guarantees full coverage (thus successful data delivery) to all terrestrial users.
Moreover, to ensure $\name_k$'s successful routing before its full deployment, each satellite $i$ in the high-altitude $\name_{k-1}$ connects to its corresponding satellites at lower altitude according to Algorithm~\ref{algo:construction}. 
In this way, if two lower-altitude satellites in $\name_{k-1}$ are unreachable due to incomplete deployment, their packets can always routed to their corresponding higher-altitude satellites in $\name_k$ for guaranteed delivery.
During this expansion, all satellites' addresses remain the same as those in $\S$\ref{sec:address:sat}.

One issue for the hierarchical deployment is to coordinate the altitudes of satellites in $\name_{k-1}$ and $\name_k$. 
Note that satellites at different altitudes have different orbital periods.
To ensure the hierarchical geographical cell divisions in $\S$\ref{sec:address}. the orbit periods of $\name_{k-1}$ and $\name_k$ should satisfy  $T_{k-1}=NT_k$.
According to the Kepler's third law $T_{k-1}^2/(R_{k-1}+R_E)^3=T_{k}^2/(R_{k}+R_E)^3$, the altitudes should satisfy 
$\frac{R_{k-1}+R_E}{R_{k}+R_E}=N^{\frac{2}{3}}$.

Although hierarchical expansion is straightforward for incremental deployment, it requires additional satellites at higher altitudes besides the original \name construction (\eg, 8+64 satellites for $\name_1$ when $N=8$). 
We next explore whether incremental expansion is possible at the same altitude. }


A straightforward way to incrementally deploy \name is to follow Algorithm~\ref{algo:construction}, \ie, launch $\name_{k-1}$ first and recursively expand it to $\name_{k}$. 
This ensures stable network topology and addresses, but can't guarantee delivery with partial coverage {\em before} all satellites are launched ($\S$\ref{sec:construction}). 

To this end, we propose an iterative incremental expansion scheme, and exploit regular satellite mobility to guarantee the data delivery. 
Instead of building $\name_k$ from $\name_{k-1}$, we incrementally deploy it {\em orbit by orbit}.
Following step 8--12 in Algorithm~\ref{algo:construction}, each time we deploy satellites that uniformly distribute on the same orbit. 
As exemplified in Figure~\ref{fig:iterative-expansion}, \name satellites in the same orbit form a moving wave of satellite sub-point trajectories that traverse the entire coverage. 
This differs from the satellite sub-points in a complete $\name_{k-1}$, which are fixed and thus limit its full coverage at lower altitude. 
As the satellite sub-point trajectories move, all users have the opportunity to send the data to the satellite sometime. 
If a packet's destination is currently outside the coverage, it can be buffered by the satellite and eventually delivered if any one of the satellite on the same orbit covers the destination. 
With more satellites deployed, more packets will be delivered via routing and less packets will be buffered. 
Similar to DTN in classical satellite networking~\cite{cerf2007delay, burleigh2003delay}, this approach is feasible when the \name users are all tolerant to data delays.

\paragraphb{Implications for ground stations:}
The ground stations can play as ``gateways'' to bridge the satellite and terrestrial networks, or download telemetry data from satellites \cite{aws-gs,ms-gs,quasar2021mobicom, adrian2020backbone}.  
It is desirable to deploy the ground stations where they can maximize its radio quality and access to more satellites.  
\name's stable satellite-to-ground mapping offers a natural solution for this.
A ground station using \name can be deployed at the intersection of the fixed satellite sub-point trajectories.
This guarantees maximal received power from the satellites (right above the ground station), and helps gain visibility to more satellites. 
\yuanjie{Highlight stability in incremental expansion: No link or address change.}

\paragraphb{Network address allocation:}
The satellites' \name addresses in $\S$\ref{sec:address:sat} are fixed and can be pre-allocated before they are launched.
For the ground users, their \name addresses in $\S$\ref{sec:address:ue} are fixed unless they move to a new geographical cell. 
To tackle it, a DHCP address server (\eg, at the ground station) can be deployed for each cell to manage the address allocation and duplication detection.
As evidenced in Figure~\ref{fig:user-address}, the DHCP server can generate a \name address by concatenating the network prefix, the geographical cell ID, and a per-user suffix that is unique inside this cell.

\paragraphb{Inter-networking to external networks:}
The discussion so far focuses on the networking inside a \name. 
In reality, both the terrestrial users and satellites need to communicate with others through the Internet. 
For compatibility with Internet routing and global reachability, we propose IPv6-based network addressing for \name. 
As shown in $\S$\ref{sec:address} and exemplified in Figure~\ref{fig:address}, 
the IPv6 prefix in Figure~\ref{fig:user-address} can be used to identify whether a packet is inside a \name. 
If so, the \name routing in $\S$\ref{sec:routing} can be applied.
Otherwise, the packet will leave \name and enter the external network.
In this case, the classical prefix match-based IP routing will be applied to the IPv6 address, thus retaining backward compatibility and global reachability. 

\paragraphb{Stability-performance tradeoff:}
\name's primarily goal is stability for a usable space-ground network.
It is possible to refine \name to balance the stability and performance.
For instance, the basic \name design only utilizes local links to for stable topology (Theorem~\ref{thm:stability}). 
It is possible to extend \name with non-local links for lower latency (first proposed in \cite{bhattacherjee2019network}), by extending the base Rosette constellation in Figure~\ref{fig:base-case} and construction in Algorithm~\ref{algo:construction} with $m$-local links.
In this case, the maximum number of hops in Theorem~\ref{thm:hop} can be reduced to $\frac{N(k+1)}{2m}$.
The cost is more link failures (Theorem~\ref{thm:stability}) and unstable routing.
We are neutral to this tradeoff and leave it for the operators to decide.

\section{Evaluation}
\label{sect:eval}


We evaluate the efficiency and overhead of \name's network structure ($\S$\ref{sec:eval:characteristics}), addressing ($\S$\ref{sec:eval:addr}), and routing  ($\S$\ref{sec:eval:routing}). 

\begin{figure}[t]
\vspace{-15mm}
\centering
\includegraphics[width=.7\columnwidth]{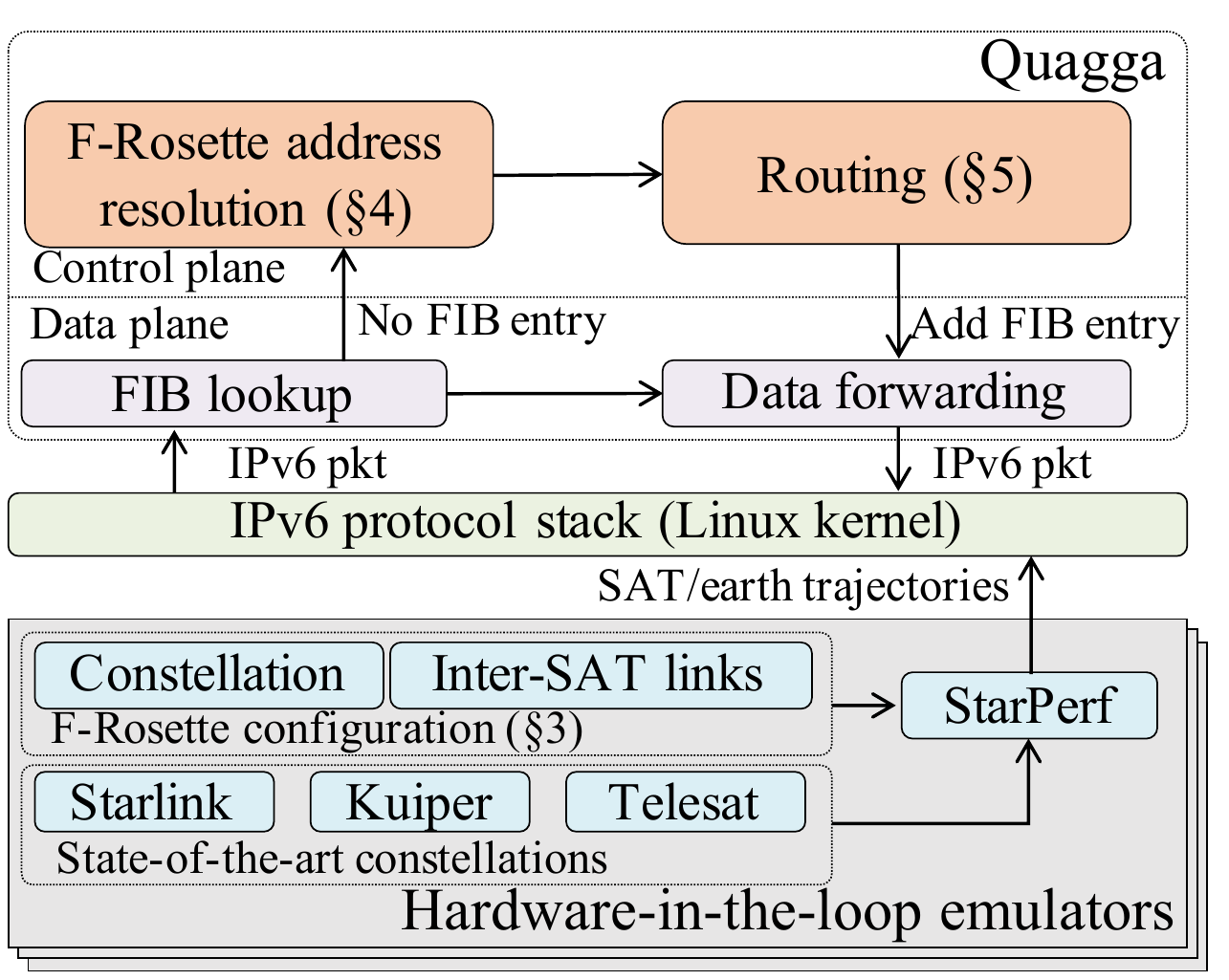}
\vspace{-2mm}
\caption{Hardware-in-the-loop experimental setup.}
\label{fig:setup}
\end{figure}
\begin{table}[t]
	\caption{
	\name under the same altitude and elevation angle of state-of-the-art mega-constellations.
	}
	\label{tab:min-satellites}
	\vspace{-2mm}
	\resizebox{1\columnwidth}{!}{{
			\begin{tabular}{|c|c|c|c|c|c|}
				\hline
				& {\bf Number of}  & {\bf Altitude}  & {\bf Inclination} &  {\bf Min. number of}\\
				& {\bf satellites} & {\bf $H$ (km)} 	& {\bf angle $\phi$} & {\bf satellites in \name}\\
				\hline
				{\bf Starlink} &1584 &550&53 & 676\\
				\hline
				{\bf Kuiper} & 1156  & 630 & 51.9& 676\\
				\hline
				{\bf TeleSat}&351 & 1015 &98.98& 225\\
				\hline
			\end{tabular}
	}}
	
\end{table}

\paragraphb{Prototype:}
We prototype an IPv6-based \name satellite protocol suite with 13K lines of C code, as shown in  Figure~\ref{fig:setup}.
We build \name on top of the standard IPv6 routing suite by extending Quagga 0.99.23 \cite{quagga} and run it in a customized Linux 3.10.0 kernel (which resembles Starlink's Linux-based small satellites \cite{starlink-cpu}).
At the control plane, we implement \name's hierarchical addresses, and embed them into the IPv6 address as shown in Figure~\ref{fig:address}.
We also implement \name's routing algorithms in $\S$\ref{sec:routing} to compute the FIB entry for a new incoming packet.
At the data plane, we reuse the standard IPv6 routing table and forwarding engine for prefix-based \name routing in $\S$\ref{sec:routing}.
At runtime, an IPv6 packet with \name address will be forwarded to the user plane.
If the routing table does not have a FIB entry for this address, it will be relayed to the control plane.
The control plane will parse the IPv6 address to get the \name address, run Algorithm~\ref{algo:single-path-routing}--\ref{algo:geo-routing} for this address, generate and install the FIB entry, and then forward the packet.
Later packets with to the same destination will be directly forwarded by routing table lookup at the data plane.

\paragraphb{Experimental setup:}
With our limited capability of launching massive satellites to space, we conduct a hardware-in-the-loop, trace-driven emulation to approximate the real space-ground networks. 
We use \textsf{StarPerf} \cite{lai2020starperf} to approximate the mega-constellations based on \name's setup in $\S$\ref{sec:construction} and real TLE orbit information of Starlink~\cite{starlink}, Kuiper~\cite{kuiper}, and Telesat~\cite{telesat} from Space-Track \cite{space-track}.
The emulator will generate the runtime satellite motions, earth's rotations, and inter-satellite link qualities (delay, loss, and capacity) based on the well-tested J4 orbit propagation model \cite{brouwer1959solution, kozai1959motion} in aerospace engineering.
For the projections in $\S$\ref{sec:challenge}, we use a ThinkStation P910 workstation (22-core 3.5GHz Intel E5-2600, 64GB DDR4) 
with \textsf{mininet} \cite{mininet} to create all virtual satellites in LEO mega-constellations. 
We also run our prototype on a testbed of 5 servers (with 2.2GHz CPU and 4GB memory usage limit to approximate the real small LEO satellites in \cite{gretok2019comparative, bhattacherjee2020orbit})
to quantify its data forwarding performance and overhead.


\paragraphb{Ethics:} This work does not raise any ethical issues. 

\subsection{Characteristics of \name Structure}
\label{sec:eval:characteristics}

\fixme{The organization seems ad-hoc, inconsistent with $\S$\ref{sec:property} and paper position (space-ground).}

We first assess the \name's network structure in $\S$\ref{sec:property}, and compare them with existing mega-constellations.

\begin{figure}[t]
\vspace{-13mm}
\includegraphics[width=.8\columnwidth]{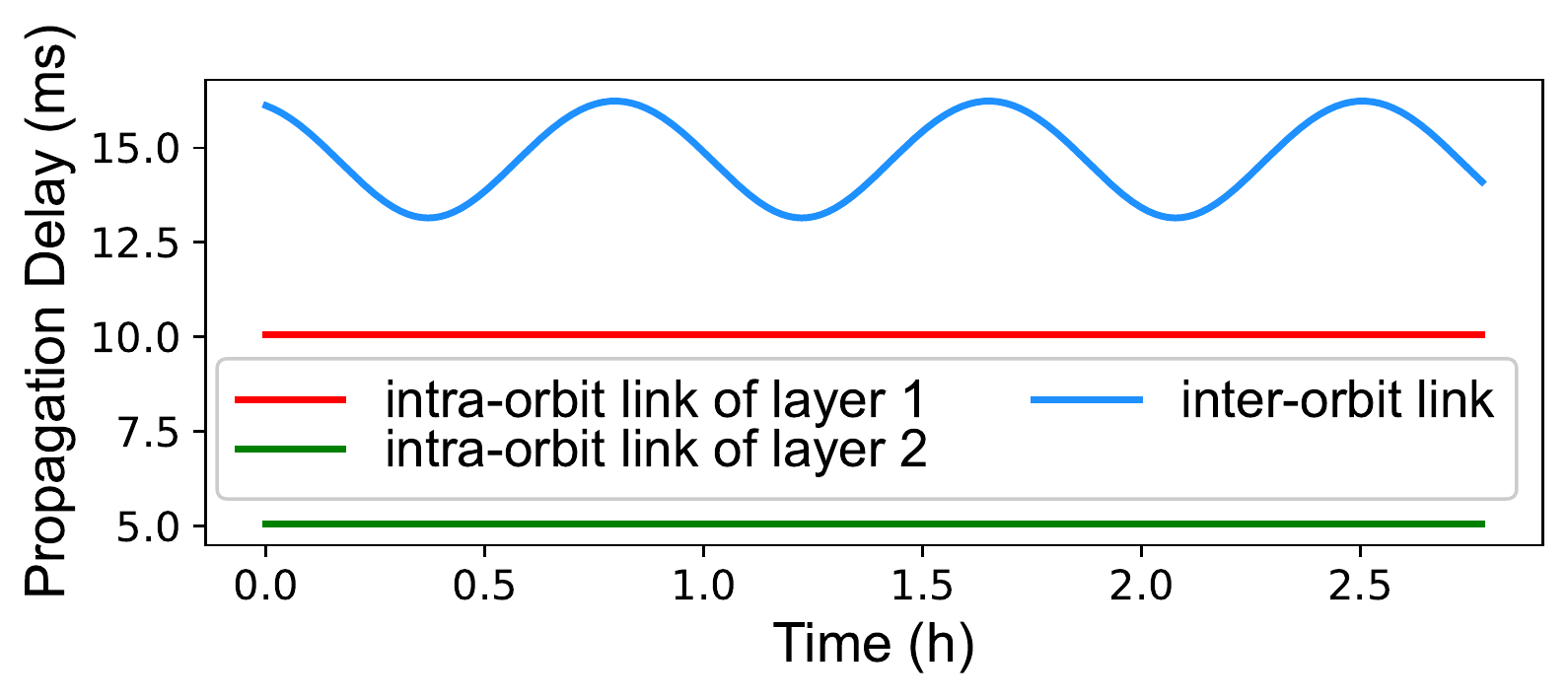}
\vspace{-2mm}
\caption{\name's inter-satellite link delay in free space under J4 orbit propagation model \cite{brouwer1959solution, kozai1959motion}.}
\label{fig:length_change}
\end{figure}

\paragraphb{Full coverage with fewer satellites:}
Table~\ref{tab:min-satellites} compares \name with state-of-the-art mega-constellations under the same altitude and elevation angle. 
It confirms \name can achieve the same coverage as the state-of-the-art's with 35.9--57.3\% fewer satellites.
This is mainly because \name is based on Rosette constellation, which typically requires less satellites than Walker-based constellations today \cite{ballard1980rosette}.
Note existing mega-constellations may still achieve full coverage with fewer satellites.
Besides, to satisfy \name's construction, the actual number of satellites in \name can be slightly more than this minimum.
This can be optimized by tuning the choices of $N$, $m$ and $k$.


\paragraphb{Stable network topology:}
\name's stability under the minimum altitude $H$ has been proven in Theorem~\ref{thm:stability}.
Table~\ref{tab:altitude} quantifies $H$ under different \name sizes.
More satellites in \name generally lead to a lower altitude, thus facilitating low latency network services.

\paragraphb{Geographical cells on the ground:} 
Table~\ref{tab:altitude} quantifies the number of geographical cells and their average sizes in \name with $N=8,m=6$.
More geographical cells can be formed by increasing the number of layers $k$, thus offering fine-grained and faster network services.
\yuanjie{To be polished with more insights and technical depth.}

\paragraphb{Propagation delay of an inter-satellite link:}
Figure~\ref{fig:length_change} exemplifies the RTT variance of links between two satellites with $N=16,k=1$ using J4 orbit perturbation \cite{brouwer1959solution, kozai1959motion}. 
For intra-orbit links, \name ensures {\em fixed} one-hop RTT because of the fixed inter-satellite distance. 
For inter-orbit links, \name confirms periodic, predictive and bounded variance of the RTT, which is feasible for low-jitter applications. 
Compared to terrestrial networks, RTT between satellites is less noisy due to the laser communication in vacuum. 

\begin{figure}[t]
\vspace{-15mm}
\subfloat[An example trace]{
\includegraphics[width=0.47\columnwidth]{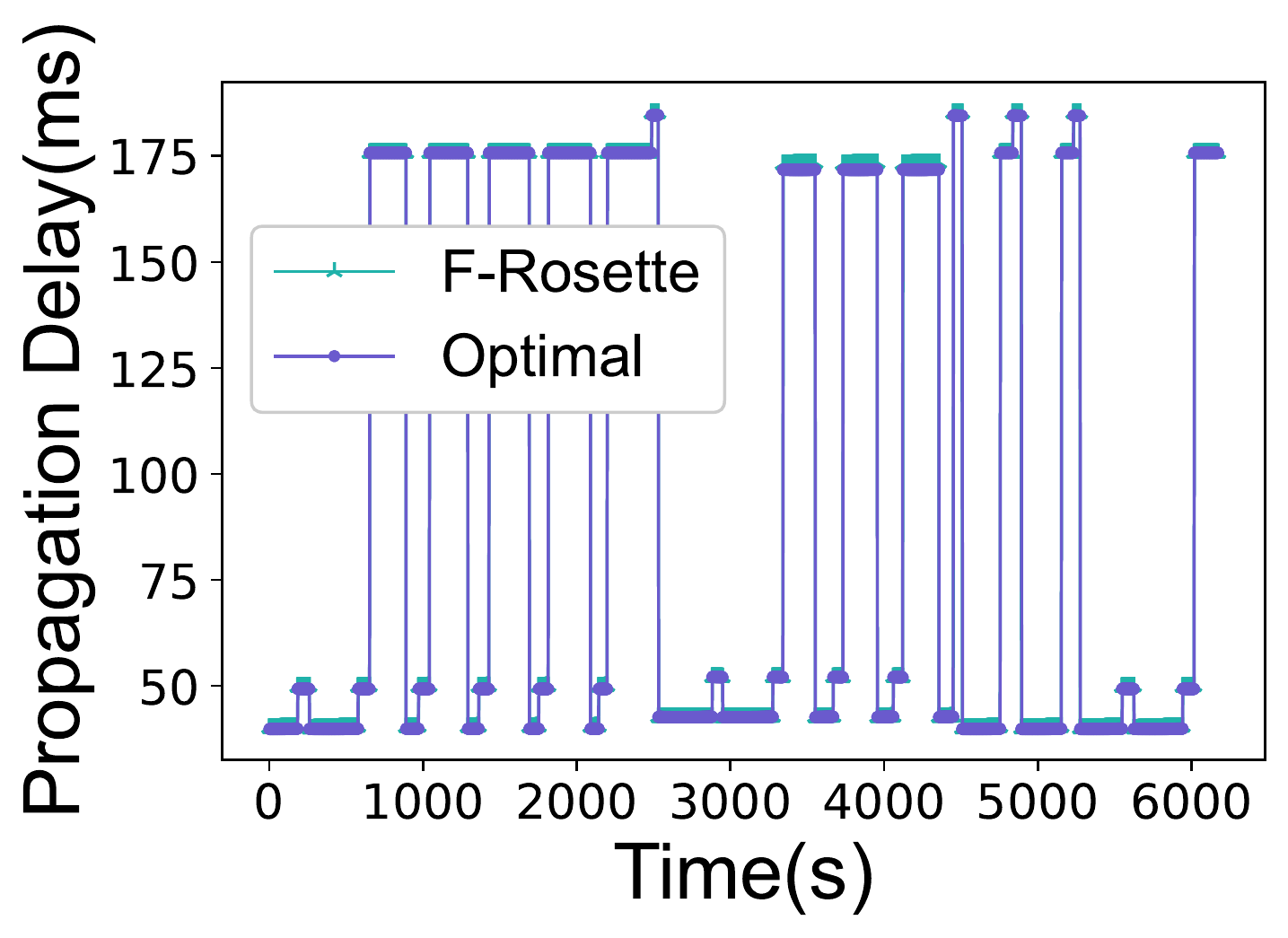}
\label{fig:delay_hop}
}
\subfloat[Closeness to optimality]{
\includegraphics[width=0.5\columnwidth]{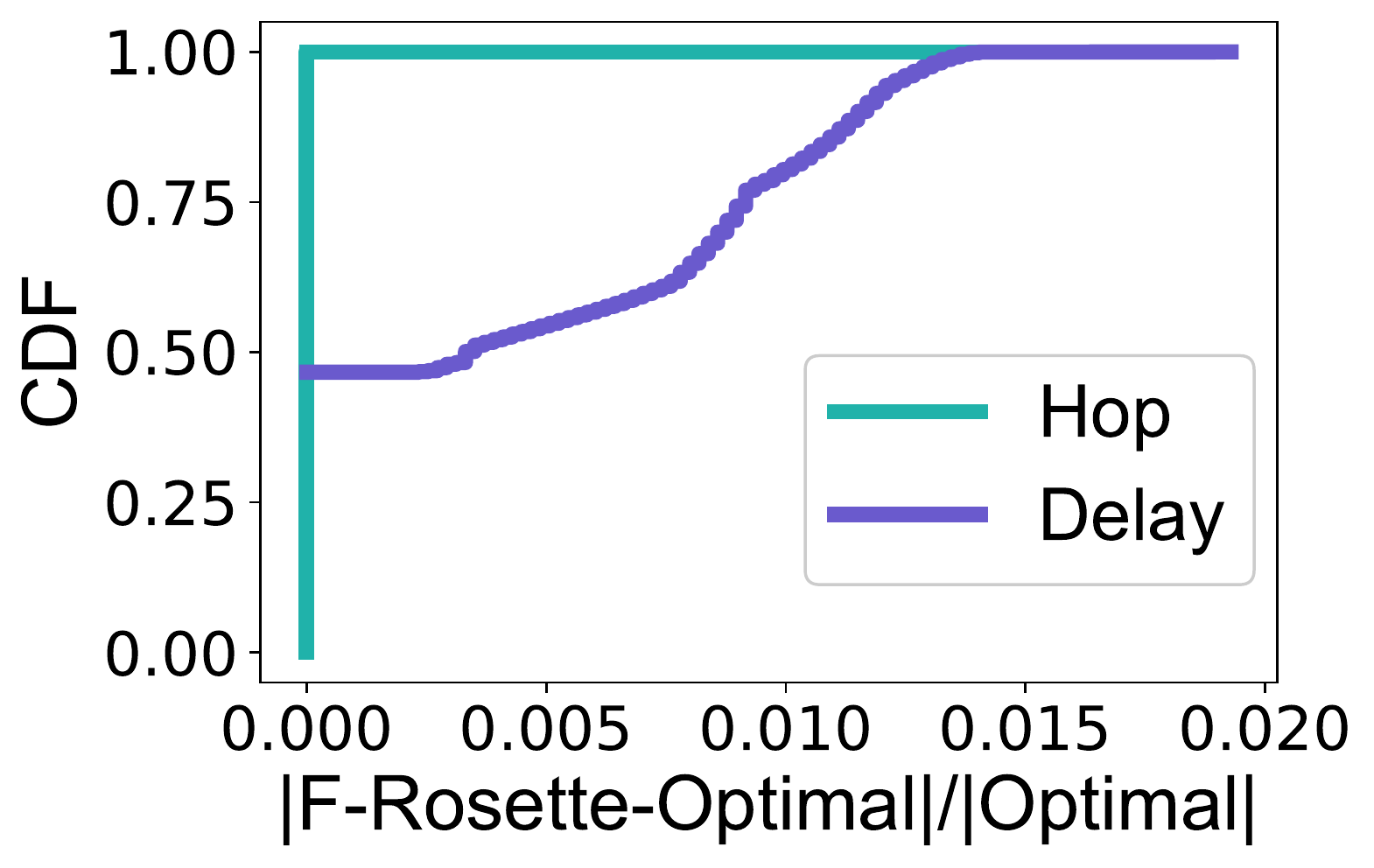}
\label{fig:delay_cdf}
}
\vspace{-2mm}
\caption{\name's near-optimal routing.
}
\label{fig:delay_hop_diff}
\end{figure}

\subsection{Network Addressing in \name}
\label{sec:eval:addr}

We next evaluate the network addressing in \name. 
It has been proved in $\S$\ref{sec:address} that \name does not incur satellite or user IP address changes in $\S$\ref{sec:challenge}. 
Here we assess the efficiency and cost of \name address in IPv6. 

\paragraphb{Addressing efficiency:}
We quantify the packet space usage by \name's address.
Table~\ref{tab:bits-16} shows the header spaces needed inside an 128-bit IPv6 address to embed \name's hierarchical address in $\S$\ref{sec:address}. 
For satellites, a \name with $N=16$ can support up to $k=8$ ($2^{32}$ satellites) with 32-bit address space. 
For ground cells, with a 32-bit address, a \name with $N=16$ can support up to $k=2$. 
More cells can be accommodated with more bits in 128-bit IPv6 address. 
Note users in the same geographical cell should be assigned different suffixes to avoid duplicate IPv6 address ($\S$\ref{sec:address:ue}).
With 32-bit suffix in an IPv6 address, \name can support up to $2^{32}$ in each geographical cell, and leaves 64 bits as IPv6 prefixes for inter-networking to external networks.

\paragraphb{Addressing overhead:}
At runtime, \name should process its address for Algorithm~\ref{algo:single-path-routing} (for satellites) or Algorithm~\ref{algo:geo-routing} (for ground users).
For the satellite address, \name performs the prefix/wildcard matching based on the standard FIB, thus retaining same cost as standard IPv6 with negligible overhead.
For ground users, \name should map its geographical cell ID to the corresponding location $(\alpha,\gamma)$.
Table~\ref{tab:prototype_overhead} confirms that the runtime table lookup time is marginal. \wei{Work with Lixin to implement the table, and evaluate its lookup time.}
Table~\ref{tab:Storage-16} shows $\leq$2MB is needed for the geographical ID-to-location table for $N=16,k=3$, 
which is acceptable even for small LEO satellites with limited storage.

\begin{table}[t]
	\caption{Header usage in \name's address ($N=16$).
	}
	\vspace{-2mm}
	\label{tab:bits-16}
	\resizebox{1\columnwidth}{!}{{
			\begin{tabular}{|c|c|c|c|c|}
				\hline
				& Num. satellites &  Bits of row &  Bits of column & Total bits\\
				\hline
				
				$\name_0$ & 16              & 4           & 4        & 8                                                               \\ \hline
				$\name_1$ & 256             & 8           & 9        & 17                                                                \\ \hline
				$\name_2$ & 4096            & 12           & 14        & 26                                                                \\ \hline
				$\name_3$ & 65536           & 16           & 19        & 35                                                                \\ \hline
			\end{tabular}
	}}
\end{table}

\subsection{Routing in \name}
\label{sec:eval:routing}

We last evaluate the routing in \name.
It has been proven in $\S$\ref{sec:routing} that \name does not incur routing re-convergence or suffer from low network usability in $\S$\ref{sec:challenge}.
So this section focuses on the routing efficiency and overhead.

\paragraphb{Routing efficiency:}
We first compare \name's routing in $\S$\ref{sec:routing} with the paths with the shortest hop counts and propagation delays.
In this experiment, we route traffic from Beijing to New York using \name with $N=16, m=2, k=1$ (256 satellites) at the altitude of 878.76km. 
We compare \name's hop counts and propagation delays with those in the shortest path.
Figure~\ref{fig:delay_hop_diff} confirms that, \name always achieves the shortest path in hop counts as proved in Theorem~\ref{thm:shortest-path}.
Due to the time-varying inter-orbit satellite link delays (Figure~\ref{fig:length_change} and Equation \ref{eqn:rtt}), \name does not always find the shortest path in propagation delays.
Even so, Figure~\ref{fig:delay_cdf} shows \name's bounded link delay variances help \name approximates the optimal delay with \add{$\leq$1.4\% difference, thus $\leq$1.62ms.} 
Note the spikes in Figure~\ref{fig:delay_cdf} arise from user's handoffs between satellites, during which the user re-associates to a closer satellite.
Despite this dynamics, \name still retains near-optimality compared to the shortest path in propagation delays.

\begin{figure}[t]
\vspace{-5mm}
\centering
\includegraphics[width=.8\columnwidth]{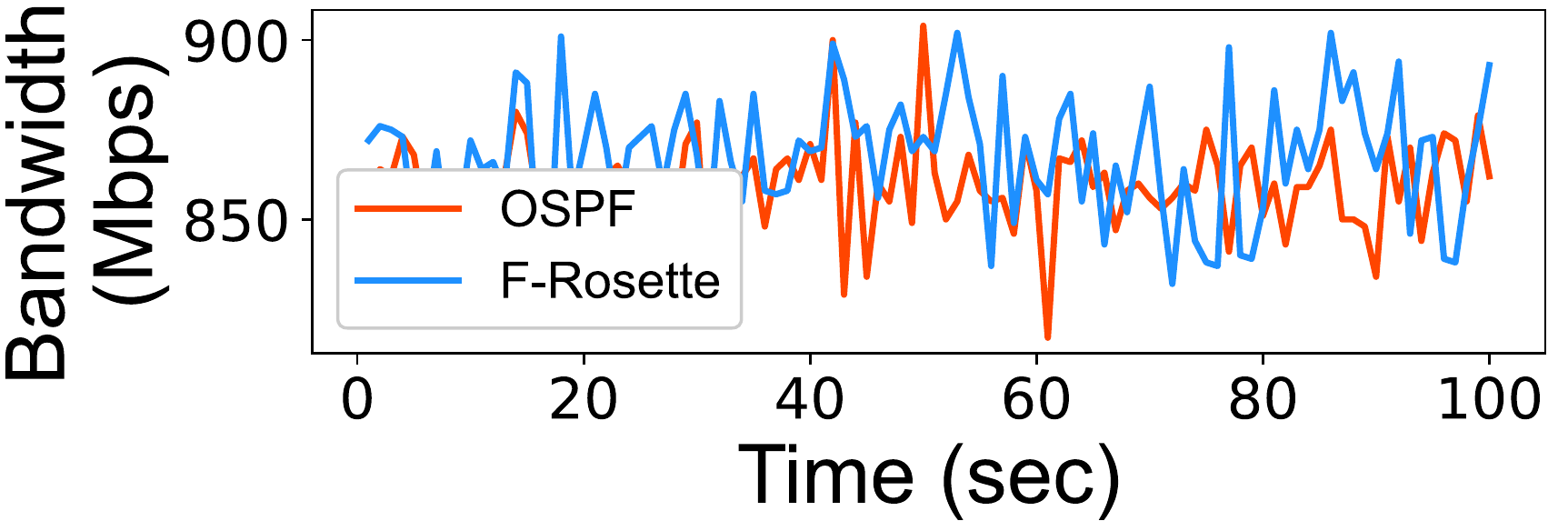}
\vspace{-2mm}
\caption{Data forwarding speed at 1Gbps links.}
\label{fig:iperf}
\end{figure}

Besides near-optimal routing path, \name's forwarding engine also retains high data throughput.
Figure~\ref{fig:iperf} shows the end-to-end \textsf{iperf} throughput in a five-server testbed running \name over 1Gbps links. \lixin{Elaborate on the setup}
It confirms \name's routing can saturate the physical bandwidth, and achieves comparable speed to the standard OSPF routing in IPv6.
Since its routing algorithm can be realized with readily-available prefix/wildcard matching today, \name achieves comparable data forwarding performance to standard IP routing. \lixin{Can we add OSPF throughput in Figure~\ref{fig:iperf} to support this point?}

\paragraphb{Routing overhead:} Since \name runs atop IPv6, it incurs marginal cost compared to state-of-the-arts.
Table~\ref{tab:prototype_overhead} shows that the processing latency of the first packet with a new destination address is $\leq$0.058ms (for FIB setup). 
Later packets to the same destination will be directly forwarded by FIB lookup without this additional processing.
Table~\ref{tab:prototype_overhead} shows $<1\%$ CPU and $\leq$1.3MB memory usage of \name under the same CPU/memory conditions as the LEO satellite's, thus affordable for resource-constrained small satellites. \lixin{Update the setup and numbers.}

\begin{table}[t]
	\caption{Storage overhead with $N=16$
	}
	\vspace{-2mm}
	\label{tab:Storage-16}
	\resizebox{1\columnwidth}{!}{{
		\begin{tabular}{|c|c|c|c|c|}
			\hline
			& $\name_0$ & $\name_1$ & $\name_2$ & $\name_3$ \\ \hline
			{\bf Memory (bytes)} & 64        & 1,984     & 61,504    & 1,906,624 \\ \hline
		\end{tabular}
	}}
\end{table}

\begin{table}[t]
\caption{System overhead of \name.}
\vspace{-2mm}
\label{tab:prototype_overhead}
\resizebox{.8\columnwidth}{!}{
\begin{tabular}{|p{5cm}|p{3cm}|}
\hline
{\bf CPU usage}  & $<1\%$ \\
\hline
{\bf Memory usage} & 1344kB\\
\hline
{\bf 1st packet's processing latency}  & 0.048--0.058ms\\
\hline
{\bf table lookup time}  & 0.006--0.008ms\\
\hline
{\bf Bandwidth} & 867.42Mbps\\
\hline
\end{tabular}
}
\end{table}

\section{Related Work}
\label{sect:related}

LEO satellite network has been actively studied for decades.
Early efforts design constellations to ensure coverage, such as Walker \cite{walker1984walker}, Rider \cite{rider1986analytic}, Rosette \cite{ballard1980rosette}, Ellipso \cite{castiel1995ellipso}, to name a few. 
LEO satellites in these constellations are standalone, without forming a network in space. 
The first LEO network in operation is Iridium \cite{tan2019new}, a 66-satellite small constellation on the polar orbit.
Since 2000s, various network designs have been proposed for Iridium, such as the location address \cite{alex2016addressing, preston2001geo, megen2001mapping, hain2002ipv6, hybridaddressing, walkeraddress}, virtual node routing \cite{ekici2000datagram, ekici2001distributed}, virtual topology routing {\cite{werner1997dynamic, chang1998fsa}}, multi-layer routing \cite{chen2005routing, lee2000satellite}, and geographical routing \cite{henderson2000distributed, lam2011geographic, kleinberg2007geographic, rao2003geographic, karp2000gpsr}. 
However, they cannot easily extend or scale to the mega-constellations as explained in $\S$\ref{sec:challenge}. 
As mega-constellations are still at early adoptions, recent studies explore their performance optimization \cite{klenze2018networking, bhattacherjee2018gearing, handley2019using} or application to broader scenarios like navigation \cite{narayana2020hummingbird}, in-orbit computing \cite{bhattacherjee2020orbit} and content delivery \cite{quasar2021mobicom, adrian2020backbone}.
\cite{bhattacherjee2019network} notices the link churns between satellites and proposes remedies, but it does not stabilize the network between space and ground.
Instead, \name takes to first step to strive for a stable space-ground network in LEO mega-constellations.

\name is inspired by many network structure designs in other domains such as data center \cite{guo2009bcube, guo2008dcell,greenberg2009vl2}, P2P \cite{stoica2001chord, ratnasamy2001scalable}, and wireless network \cite{grossglauser2002mobility, diggavi2002even, zhou2012mirror}. 
It is based on the Rosette constellation \cite{ballard1980rosette}, but generalizes it with a recursive structure and inter-satellite links. 
Its network addressing is inspired by classical geographical information systems like Google S2 \cite{google-s2} and Uber H3 \cite{uber-h3}, but customized for space-ground networking.
\name's geographical-to-topological routing embedding resembles greedy embedding \cite{kleinberg2007geographic, lam2012geographic}, but works in a different domain of space-ground network.

\section{Conclusion}
\label{sect:concl}

This work devises \name, a stable space-ground network structure in LEO mega-constellation.
\name strives for stable network topology, address, and routing under dynamic many-to-many space-ground mapping in high mobility.
\name's recursive structure over the Rosette constellation guarantees provably stable network topology. 
Its hierarchical, geographical addressing embedded prevents the frequent address change for its users.
\name does not incur global routing re-convergence in satellite mobility or earth's rotations.
To route traffic between users, \name performs efficient geographical routing by embedding it into the stable topological routing between satellites.
\name is compatible with IPv6 and incrementally deployable.

\name is an initial effort to strive for a stable space-ground network at scale.
Compared to the terrestrial network nodes, satellites are much harder to modify or upgrade once they are launched to space. 
Therefore, although the mega-constellations are still at early deployments, it is necessary to systematically design the network before it is too late to change.
We hope \name could call for the community's attention and stimulate more research efforts in this area. 


\bibliographystyle{unsrt}
\bibliography{bib/extreme-mobility,bib/proposal,bib/satellite,bib/standard,bib/wing,bib/cellular}

\appendix

\section{Proof of Theorem~\ref{thm:coverage}}
\label{proof:coverage}

\begin{proof}
	
	By construction, \name guarantees the full coverage by following the condition for Rosette constellation \cite{ballard1980rosette}.
	With $N^{k+1}$ satellites in a $\name_k$, Equation \ref{eqn:per-sate-coverage} in $\S$\ref{sec:rosette-primer} has shown it suffices for each satellite with coverage $R=\sec^{-1}\left(\sqrt{3}\tan{\left(\frac{\pi}{6}\frac{N^{k+1}}{N^{k+1}-2}\right)}\right)$ to guarantee full coverage. 
	Given this per-satellite coverage $R$ and elevation angle $\phi$, the minimum altitude that a satellite can achieve this satellite is given by Equation \ref{eqn:altitude}.
	Therefore, we have $H_{min}=R_E\left(\frac{1}{\cos R-\sin R\tan\phi}-1\right)$.

\end{proof}

\section{On-Demand \name Size Selection}
\label{on-demand-size}

A satellite operator can choose the proper \name size that can ensure full coverage {\em and} meet its network latency demands.
Algorithm~\ref{algo:layout} illustrates the size selection based on the ground-to-space round trip latency demand \textsf{RTT}, and the minimum elevation angle $\phi$ (constrained by the satellite and device hardware's capability).
The operator can first specify the desired altitude of the satellites $H=c\cdot\textsf{RTT}/2$, where $c$ is the light speed through the atmosphere. 
With the minimum elevation angle $\phi$, one we derive the minimal per-satellite coverage $R_{min}$ in the unit of great circle range:
$\tan\phi=\frac{\cos R_{min}-R_E/(R_E+H)}{\sin R_{min}}$.
With $R_{min}$ the operator can derive the minimal number of satellites needed for a Rosette constellation to guarantee full coverage:
$\sec R_{min}=\sqrt{3}\tan{\left(\frac{\pi}{6}\frac{N_{min}}{N_{min}-2}\right)}$.
Then given the $\name_0$ parameter $N$, the operator can choose the level of \name as $k=\lceil{\log_N{N_{min}}}\rceil-1$.

\begin{algorithm}[t]
   \centering
        \begin{algorithmic}[1]
    	\Require{Ground-to-satellite round-trip demand \textsf{RTT}, minimum elevation angle $\phi$} 
	\Ensure{The minimum number of layers $k$ and satellites $N_k$  for \name to guarantee full coverage}
	\State Compute satellite altitude $H=c\cdot\textsf{RTT}/2$; 
	\State Find each satellite's minimum coverage $R_{min}$ in the unit of great circle range: 
	$$\tan\phi=\frac{\cos R_{min}-R_E/(R_E+H)}{\sin R_{min}}$$
	
	\State Decide the minimum number of satellites to guarantee full coverage:
	
	$$\sec R_{min}=\sqrt{3}\tan{\left(\frac{\pi}{6}\frac{N_{min}}{N_{min}-2}\right)}$$
	
	\State Decide the minimum number of layers $k$ for \name:
	
	 $$k=\lceil{\log_N{N_{min}}}\rceil-1$$

	\State Decide the minimum number of satellites in \name: $N_k=N^{k+1}$;

    \end{algorithmic}
    \caption{On-demand \name size selection.} 
    \label{algo:layout}
\end{algorithm}

\section{Proof of Theorem~\ref{thm:stability}}
\label{proof:stability}

\begin{proof}
To ensure the inter-satellite links are always on, the altitudes of the satellites should be sufficiently high for mutual visibility.
To derive the minimum altitude, we first compute the great circle range $r_{ij}$ of satellite $i$ and $j$ from Equation~\ref{eqn:rij}.
When $r_{ij}$ is maximized, the distance between the two satellites reaches the maximum, which requires the highest altitude for mutual visibility. 
Therefore, if we want to prevent the link from passing through the earth, the condition that needs to be met is $(H+R_E)>\frac{R_E}{cos\frac{r_{max}}{2}}$, where 
\begin{align*}
	\sin^2(r_{max}) &= \cos^4(\beta/2)\sin^2m(m+1)(\pi/N)\\
	&+ 2\sin^2(\beta/2)\cos^2(\beta/2)\sin^2m^2(\pi/N)\\
	&+ \sin^4(\beta/2)\sin^2m(m-1)(\pi/N)\\
	&+ 2\sin^2(\beta/2)\cos^2(\beta/2)\sin^2m(\pi/N)\\ 
\end{align*}
Moreover, to guarantee full coverage, we require $H>H_{min}$. This ends up with $H>\max\{(\frac{1}{cos\frac{r_{max}}{2}}-1)R_E,H_{min}\}$.
\end{proof}

\section{Proof of Theorem~\ref{thm:hierarchy}}
\label{proof:hierarchy}

\begin{proof}
We first note that each \name cell is a unique quadrilateral bounded by four satellite sub-point trajectories (two pairs of parallel spherical edges).
To prove Theorem~\ref{thm:hierarchy}, it is sufficient to prove that when constructing $\name_k$ out of $\name_{k-1}$, each $\name_{k-1}$ cell will be intersected by $N-2$ new satellite sub-point trajectories parallel to two of its edges, and another $N-2$ satellite sub-point trajectories parallel to the other two of its edges.
In this way, this $\name_{k-1}$ cell is divided into $N^2$ cells in $\name_k$.
To prove this, consider a $\name_{k-1}$ cell bounded by the satellite sub-point trajectories of satellite $S_i$, $S_{i+1}$, $S_j$ and $S_{j+1}$, as shown in Figure~\ref{fig:proof-hierarchy}. Recall the satellite sub-point trajectory for a satellite $S_i$ in $\name_{k-1}$ in Equation~\ref{eqn:phi}--\ref{eqn:lambda}.
To construct $\name_{k-1}$ out of $\name_k$, each satellite $S_i$ in $\name_{k-1}$ is replicated $N$ times, each being shifted by a uniform time interval $t_k=\frac{T_s}{N^k}, k=0,1,2,...,N-1$.
Note $t_0=0$ overlaps with $S_i$'s satellite sub-point trajectory.
For each of the remaining replicated satellite $S_i^k$, its satellite sub-point trajectory is thus shifted by $t_k$ (dotted lines in Figure~\ref{fig:proof-hierarchy}), which is between the satellite sub-point trajectories of $S_i$ and $S_{i+1}$.
Similar divisions apply to satellite $S_j$ and $S_{j+1}$.
So we conclude that this $\name_{k-1}$ cell is divided into $N^2$ $\name_k$ cells.
\add{However, some $\name_{k}$ cells near the polar regions do not belong to any  $\name_{k-1}$ cells , so we artificially add two latitude lines equal to the inclination angle to form some  $\name_{k-1}$ cells . These cells will be divided into  $\frac{N^2+N}{2}$  $\name_{k}$ cells . In order to satisfy this theorem , we will calculate some small cells repeatedly in the big cell. }
%
\end{proof}

\section{Proof of Theorem~\ref{thm:fib}}
\label{proof:fib}

\begin{proof}
Consider a satellite $S=s_0.s_1....s_{k}$ in a $\name_k$.
At each layer $j\in[0,k-1]$, $S$ belongs to a ring topology of $N$ satellites according to \name's construction. 
According to Algorithm~\ref{algo:single-path-routing}, these $N$ satellites are divided into 2 groups based on $S$'s shortest path to them: one group is routed via clockwise routing, and the other group is routed via counter-clockwise routing.
All satellites in the same group share the same next hop at $S$, so they can be aggregated in the FIB.
Therefore, the number of FIB entries $S$ needs at layer $j$ is the number of prefixes that $(s_j,s_j+\frac{N}{2}\mod N]$ and $(s_j+\frac{N}{2}\mod N,s_j]$ can be aggregated into, which are both $\lceil\log\frac{N}{2}\rceil$.
So $S$ needs $2\lceil\log\frac{N}{2}\rceil$ FIB entries for layer $j$.
Since there are $k+1$ layers in a $\name_k$, each satellite needs $2(k+1)\lceil\log\frac{N}{2}\rceil$ FIB entries.
\end{proof}

\section{Proof of Lemma~\ref{lem:cells}}
\label{proof:cells}

\begin{proof}

\begin{figure}[t]
	\includegraphics[width=\columnwidth]{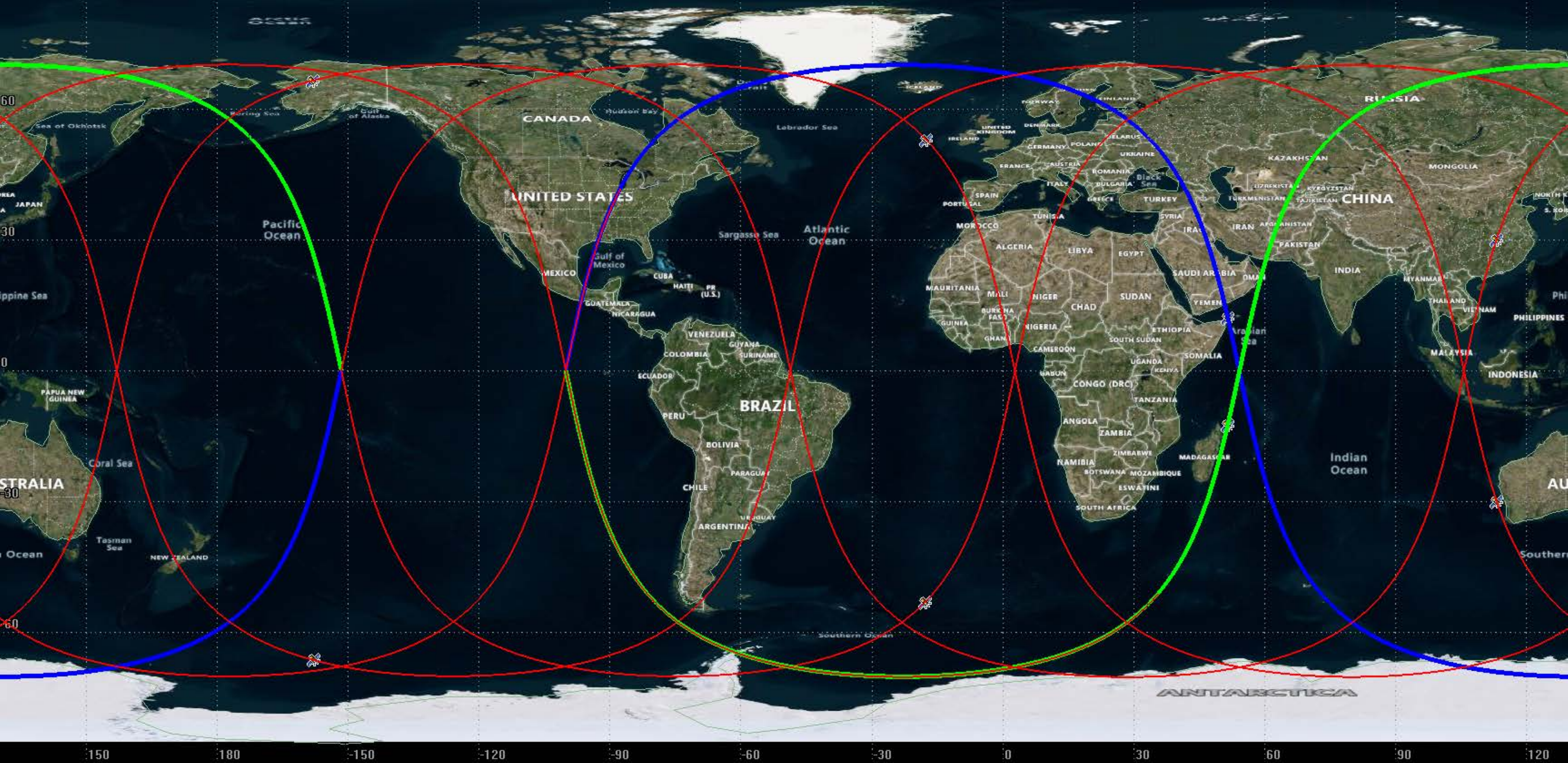}
	\caption{cells number in \name.}
	\label{fig:cells-number-proof}
\end{figure}

We show that a $\name_0$ has $(N-m)^2$ cells.
Then according to Theorem~\ref{thm:hierarchy}, a $\name_k$ has $(N-m)^2N^{2k}$ cells.
First note \name's repeat ground-track orbits dedicate $T_E=(N-m)T$,
\ie, after an orbital period $T$, the earth rotates $\frac{1}{N-m}$ of the great circle. 
So it takes $(N-m)T$ for a satellite in $\name_0$ to revisit the place it passed before. 
Based on the repetitions of the orbit period, a satellite's sub-point trajectory can be divided in to $N-m$ parts $\{C_i, i=1,2,...,N-m\}$.
As Figure~\ref{fig:cells-number-proof} exemplifies, the blue curve is one of $C$, and curves will intersect to form cells. 
We can uniquely identify a cell with its left vertex since the number of the vertexes is equal to the number of the cells (Euler theorem).
Because of the earth's rotation, there are 1 curve that do not intersect with $C_i$ (green curve in Figure~\ref{fig:cells-number-proof}).
Besides, $C_i$ has $2(N-m-2)$ intersections with others (excluding itself).
So there are $(N-m)(N-m-2)$ intersections in total.
In addition, with the maximum latitudes a $\name_0$ can cover (constrained by the satellites' inclination angle $\beta$), 
two more rows exist at these latitudes and add the total number of cells to $(N-m)^2$.
So according to Theorem~\ref{thm:hierarchy}, a $\name_k$ has $(N-m)^2N^{2k}$ cells .
\end{proof}

\section{Proof of Theorem~\ref{thm:multipath}}
\label{proof:multipath}

\begin{proof}
We prove this theorem by recursion. 
When $k=0$, there are always $2(k+1)=2$ disjoint paths between any two satellites (one clockwise, the other counter-clockwise as exemplified in Figure~\ref{fig:base-case}).
Assume there are $2(k+1)$ disjoint paths for a $\name_{k}$.
Now consider a $\name_{k+1}$ and two satellites $S=s_0.s_1....s_k.s_{k+1}$ and $D=d_0.d_1....d_k.d_{k+1}$.
The following $k+2$ routes enumerate all disjoint paths from $S$ to $D$:

$$s_0.s_1....s_k.s_{k+1}\rightarrow d_0.s_1....s_k.s_{k+1}\rightarrow d_0.d_1....d_k.d_{k+1}$$
$$s_0.s_1....s_k.s_{k+1}\rightarrow s_0.d_1....s_k.s_{k+1}\rightarrow d_0.d_1....d_k.d_{k+1}$$
$$...$$
$$s_0.s_1....s_k.s_{k+1}\rightarrow s_0.s_1....s_k.d_{k+1}\rightarrow d_0.d_1....d_k.d_{k+1}$$
For each route, the first hop has 2 disjoint paths (one clockwise, the other counter-clockwise), and the second hop has $2(k+1)$ disjoint paths.
So each route forms $\min(2,2(k+1))=2$ disjoinZt paths.
Since each two routes above do not overlap, we end up with $2(k+2)$ disjoint routes from $S$ to $D$ and thus conclude this proof by recursion.
\end{proof}

\section{Geographical Cell ID-to-Location Table}
\label{id-to-location-table}

A key prerequisite in Algorithm~\ref{algo:geo-routing} is to decide the destination cell's geographical location $(\alpha, \gamma)$ given its ID $c_0.c_1....c_k$ for each packet.
This section describes how to accelerate this operation via table lookup.

\paragraphb{Table structure:} 
Figure~\ref{fig:geo-routing} illustrates the mapping table from geographical cell ID $c_0.c_1....c_k$ to its location $(\alpha, \gamma)$ (defined as its left vertex).
As we will see, this table only needs to store $\alpha$ and $\gamma$ can be derived from $\alpha$.
A $\name_k$ constructs one table for level 0, and $\frac{(N-m)N^{i-1}}{2}$ tables for level $i=1,2,...k$.
Each table contains two columns, indicating the row index $c_j.\textsf{row}_j$ and the corresponding coordinate of the {\em first} cell in each row $\alpha_0$. \add{The row of the previous level plus its index multiplied by N represents the index of the next level table.}
For any other cell $i$ in this row, its coordinate can be derived as $\alpha_j=\alpha_0+\frac{2\pi j}{\left( N-m \right) \cdot N^k}$ according to \name's construction.
Each cell $i$'s phase shift $\gamma_j\equiv (N-m)\alpha_0$.

\paragraphb{Table pre-computation:}
To pre-compute the mapping table, the key step is to learn $\alpha_0$ for each row.
To derive $\alpha_0$, we have two observations.
First, all cells in \name are formed by the satellite sub-point trajectories, so $\alpha_0$ can be derived from Equation~\ref{eqn:phi}--\ref{eqn:lambda}.
Second, \name guarantees all satellites traverse the same stable sub-point trajectories and thus cells ($\S$\ref{sec:rosette-primer}--\ref{sec:construction}).
So $\alpha_0$ can be learned by tracking {\em any} satellite in \name. 
For simplicity, we derive $\alpha_0$ from the first satellite (0.0...0) in \name as $\alpha_0=\omega_Et$.
According to Equation~\ref{eqn:lambda}, its sub-point trajectory is:
$$\tan \left( \lambda +\omega _Et \right) =\cos \beta \cdot \tan \left( m\omega _Et \right)$$
From $t$ to $t+T$, this satellite passes through all rows of cells with an increment of longitude $\Delta\lambda/2$, where $\varDelta \lambda =\frac{2\pi}{\left( N-m \right) \cdot N^k}$ is the longitude difference between adjacent cells.
So given $\Delta\lambda/2$ and $T$, we numerically search all the time $t$ that satisfies the above equation, and then get $\alpha_0=\omega_Et$.

\begin{algorithm}[t]
		\centering
\begin{algorithmic}[1]
	\Require{Cell ID $c_0.c_1.....c_k$. $N$, $m$, $k$, table lists of $\alpha_0$ $\textsf{Table}_{i},i=0,1,...,k$,each $\textsf{Table}_{i}$ is a list of tables, target level of cell $k_0$}
	\Ensure{Geographical location $(\alpha,\gamma)$ }
	\State $\textsf{index}=0$; \textcolor{gray}{\Comment{{\em\tiny Index in table list, indicates which table to look up}}}
	\State $\Delta \alpha=c_{0}.\textsf{col}\cdot \frac{2\pi}{N(N-m)}$;
	\For{i=1 to $k_0$}
	\State $\textsf{index}=\textsf{index}\cdot N+c_{i-1}.\textsf{row}$;
	\State $\Delta \alpha=\Delta \alpha+c_{i}.\textsf{col}\cdot \frac{2\pi}{(N-m)N^i}$; 
	\EndFor
	\State $T=Table_{k_0}[\textsf{index}]$; \textcolor{gray}{\Comment{{\em\tiny The table to look up}}}
	\State $\alpha_0=T[c_{k_0}.\textsf{row}]$; \textcolor{gray}{\Comment{{\em\tiny Get the $\alpha_0$ from table $T$}}} 
	\State $\alpha=\alpha_0+ \Delta \alpha$; $\gamma=(N-m)\alpha_0$;
%
		\State \Return $(\alpha,\gamma)$;
		\end{algorithmic}
	\caption{Mapping geographical cell ID to its location in \name.}
	\label{algo:cell_location}
\end{algorithm}

\paragraphb{Hierarchical table lookup:}
To learn the location $(\alpha,\gamma)$ from a geographical cell ID $c_0.c_1...c_k$, a satellite hierarchically searches the mapping table level-by-level.
Algorithm~\ref{algo:cell_location} illustrates the table lookup procedure.
\name maintains the accumulative relative position $\Delta\alpha$ to the first cell in each row $\alpha_0$.
Starting from the first level, \name iteratively updates $\Delta\alpha$ and decides the table it should search in the next level.
Then this cell's location is decided as $\alpha=\alpha_0+\Delta\alpha, \gamma=(N-m)\alpha_0$.

\paragraphb{Storage and lookup cost:}
Both are modest for the small LEO satellites.
With one entry per geographical row, a satellite should store $\frac{(N-m)N^k}{2}$ entries in the table (\eg, $\approx$2MB as shown in Table~\ref{tab:Storage-16}). 
The table lookup traverses one row per \name level, resulting in a $O\left( k+1 \right)$ lookup complexity.

\end{document}